\documentclass[11pt,letterpaper]{article}
\usepackage[utf8]{inputenc}
\usepackage{amsmath,amssymb,amsthm}
\usepackage{xcolor}
\usepackage[hyphens]{url}
\usepackage{hyperref}
\usepackage{listings}
\usepackage{microtype}
\usepackage{geometry}
\usepackage{booktabs}
\usepackage{cite}
\usepackage{tabularx}
\usepackage{longtable}
\newtheorem{theorem}{Theorem}

\lstdefinelanguage{VeGo}{
  morekeywords={package,import,func,return,type,struct,interface,for,if,else,var,nil,true,false,bool,int,string},
  sensitive=true,
  morecomment=[l]{//},
  morecomment=[s]{/*}{*/},
  morecomment=[l]{//@},
  morestring=[b]"
}

\title{\textbf{VeGo: A Lightweight Deductive Verification System for Go Designed for Educational Transparency}}

\author{
  \textbf{Tina Masoudi} \quad \textbf{Christopher Dutchyn} \\
  Department of Computer Science \\
  University of Saskatchewan \\
  Saskatoon, SK, Canada \\
  \texttt{uka071@usask.ca} \quad \texttt{cjd032@usask.ca}
}

\date{October 2026 (VeGo v0.9.4)}

\begin{document}

\maketitle

\begin{abstract}
Integrating formal program verification into undergraduate computer science education is historically hindered by cognitive friction: students must master mathematical logic alongside specialized, non-standard verification languages (such as Dafny) or heavyweight interactive provers. In this paper, we present \textbf{VeGo} (Verified Go), a deductive formal verification system that brings Hoare-style verification directly into mainstream Go source code. Specifications---including preconditions, postconditions, loop invariants, integer variants, recursive measures, and framing assertions---are embedded within non-intrusive \texttt{//@} comment annotations, allowing annotated programs to compile and execute natively with standard Go toolchains.

VeGo translates procedural Go into Static Single Assignment (SSA) form, generates verification conditions via weakest precondition calculus, and discharges obligations using an in-process native Go deduction engine without external binary dependencies. In default gradual mode, unproven executable assertions are residualized into dynamic runtime checks under an assume-after-assert discipline, while unproven non-executable assertions remain static diagnostics. To guard against underconstrained specifications, we introduce \textbf{Anti-VeGo}, an adversarial counter-synthesis engine that probes contracts against degenerate behavioral archetypes. We state and sketch the soundness of a core procedural fragment with explicit runtime error states, delineate its explicit Trusted Computing Base, and evaluate VeGo across a comprehensive benchmark corpus of 108 programs (89 textbook algorithms and data structures from CLRS, Kleinberg--Tardos, Okasaki, and Knuth's TAOCP, alongside 19 modular, concurrent, and curricular programs). Under strict verification mode, 27 of 108 benchmarks fully discharge all verification conditions statically with zero unproven obligations (comprising 17 textbook algorithms and data structures, and 10 modular domain and regression benchmarks). Under default gradual mode, all 108 benchmarks execute with dynamic safety residualization, statically proving 66.4\% of bounds safety obligations (modulo integer overflow) and 91.5\% of termination obligations. Furthermore, Anti-VeGo identifies specification loopholes in 47.1\% of evaluated corpus functions (147 of 312), exposing underspecified contracts to guide student contract strengthening. Although the toolchain has been exercised in classroom instruction (a pilot in CMPT 263), a controlled empirical study of student learning outcomes is framed as future work.
\end{abstract}

\section{Introduction}
Formal software verification is essential for constructing high-assurance computing systems. However, incorporating formal methods into the core undergraduate computer science curriculum remains challenging. Instructors typically face a sharp pedagogical trade-off: either present formal logic abstractly using manual paper-and-pencil proofs without tool assistance, or require students to learn a specialized specification language (e.g., Dafny~\cite{leino2010dafny}, Why3~\cite{filliatre2013why3}, or Coq/Isabelle) that is disconnected from the mainstream imperative programming languages students use in their software engineering projects.

The pedagogical foundation for integrating software specification directly into imperative programming instruction was articulated by R.D.\ Tennent in \emph{Specifying Software}~\cite{tennent2002specifying}. Tennent demonstrated that rigorous reasoning about correctness belongs alongside imperative code, pioneering comment-embedded annotations (defining an empty \texttt{\#define ASSERT(P)} macro so that assertions act as non-executing comments rather than runtime checks), C-like bounded quantifiers (\texttt{ForAll (i=a; i<b) P}), and slice-range assertions (\texttt{A[a:b-1]}). While Tennent's curriculum relied on manual proof tableaux without automated tool assistance (and rejected Java for teaching specifications due to reference aliasing and object-oriented overhead, anticipating our language rationale), VeGo modernizes this pedagogical lineage by pairing comment-based specifications with automated SSA verification condition generation and in-situ deduction.

To realize this pedagogical vision, we developed \textbf{VeGo} (Verified Go). VeGo enables students and instructors to write formal specifications directly inside standard Go source files using special comment annotations (\texttt{//@}), drawing upon the established tradition of behavioral interface specification languages such as JML~\cite{leavens2006jml} and ACSL~\cite{baudin2008acsl}. Because specifications reside in comments, annotated \texttt{.vgo} files that omit syntactic sugar (such as \texttt{while} loops) remain valid Go syntax that, renamed or transpiled to \texttt{.go}, compile and execute natively using the standard Go compiler (\texttt{gc}), while simultaneously undergoing static deductive verification via the \texttt{vegop} verifier and Visual Studio Code Language Server Protocol (LSP) extension.

By \emph{educational transparency}, we denote three core architectural design principles: (1) \emph{syntactic transparency}, embedding specifications directly into standard Go comments without inventing a specialized host dialect; (2) \emph{algorithmic transparency}, implementing the deduction solver as a pure Go in-process engine whose intermediate SSA forms and proof goals are directly inspectable by students without external solver binary dependencies; and (3) \emph{diagnostic transparency}, delivering immediate IDE-level feedback, runtime residual assertions, and adversarial counterexamples that explain \emph{why} a specification or proof fails.

In this work, we make the following contributions:
\begin{enumerate}
    \item \textbf{A Deductive Verifier for Go}: We design and implement a deductive formal verification system tailored to a well-defined procedural subset of Go, embedding preconditions, postconditions, loop invariants, and well-founded recursive measures in source-level comments.
    \item \textbf{Native SSA Verification Pipeline}: We construct an automated verification pipeline that translates Go control flow into Static Single Assignment (SSA) form, generates verification conditions ($VC$) via weakest precondition ($wp$) calculus, and discharges them using a native Go deduction engine (equational simplification, modus ponens with recursive predicate unfolding, bounded quantifier instantiation, and Fourier--Motzkin linear inequality elimination) without external runtime binary dependencies. In default gradual mode, undischarged executable assertions are automatically residualized into dynamic runtime checks under an assume-after-assert discipline~\cite{bader2018gradual,wise2020gradual,zimmerman2024sound,divincenzo2024gradualc0}.
    \item \textbf{Semantic Model and Soundness Characterization}: We define the supported Go fragment and its semantic model with explicit runtime error states ($\bot_{\text{err}}$), state and sketch a soundness theorem for the core procedural fragment, and delineate the explicit Trusted Computing Base (TCB).
    \item \textbf{Contract Precision and Vacuity Auditing}: We provide automated contract precision audits to detect insufficient preconditions ($P_{\text{decl}} \implies wp(C, \text{safe})$) and heuristic pattern checks for under-specified postconditions.
    \item \textbf{Anti-VeGo Adversarial Counter-Synthesis}: We introduce an adversarial testing engine inspired by mutation analysis and IronSpec~\cite{goldweber2024ironspec} that inverts the verification query, synthesizing degenerate implementations and applying systematic mutations to expose specification loopholes across a six-tier strength hierarchy.
    \item \textbf{Pedagogical Resource Coverage and Classroom Exercise}: We evaluate VeGo across 108 benchmark programs (89 classical textbook algorithms and data structures drawn from CLRS~\cite{cormen2009introduction}, Kleinberg--Tardos~\cite{kleinberg2006algorithm}, Okasaki~\cite{okasaki1998purely}, and Knuth's TAOCP~\cite{knuth1997taocp,knuth1998vol3,knuth2011vol4a}, alongside 19 modular, concurrent, and curricular routines), demonstrating that the technique and tools apply to substantial pedagogical resources. We report on exercising the system in classroom settings and frame a comprehensive empirical study of student learning outcomes as future work.
\end{enumerate}

The remainder of this paper is organized as follows. Section~\ref{sec:language} details the language selection rationale and defines the supported Go fragment and semantic model. Section~\ref{sec:ai} examines formal verification as a guardrail in AI-assisted code generation. Section~\ref{sec:specs} presents the VeGo specification language and modular standard library interface contracts. Section~\ref{sec:architecture} describes the system architecture and verification pipeline. Section~\ref{sec:soundness} details the core soundness guarantees, TCB, and architectural dogfooding. Section~\ref{sec:antivego} presents the Anti-VeGo adversarial engine. Section~\ref{sec:evaluation} evaluates algorithm case studies, quantitative benchmark performance, and our 4-pillar empirical assurance framework. Section~\ref{sec:related} discusses related work, Section~\ref{sec:classroom} outlines pedagogical resource application, classroom exercise, and future empirical evaluation, Section~\ref{sec:limitations} addresses limitations and threats to validity, and Section~\ref{sec:conclusion} concludes.

\section{Language Selection Rationale and Semantic Model}
\label{sec:language}

\subsection{Why Go for Educational Verification?}
A central design decision in building VeGo was selecting an imperative programming language suitable for undergraduate instruction:

\begin{enumerate}
    \item \textbf{Purely Functional Languages (Haskell, OCaml, SML)}: While functional languages excel at equational reasoning and immutability, core computer science curricula require students to master mutable state transformations, in-place array updates, and iterative algorithms. Retaining an underlying imperative language is essential for algorithmic breadth.
    \item \textbf{C}: Early C lacked native boolean types, and C programs are prone to unchecked pointer arithmetic, memory corruption, and undefined behavior. While modern C (C99/C23) provides standardized booleans, manual memory management and pointer pitfalls impose substantial cognitive overhead on students learning formal reasoning for the first time.
    \item \textbf{Java}: Traditional Java requires extensive object-oriented boilerplate (enclosing classes, package hierarchies, and wrapper objects). While Java benefits from mature specification systems like JML~\cite{leavens2006jml} and KeY~\cite{ahrendt2016key}, formulating simple contracts requires navigating class hierarchies and reference types.
    \item \textbf{Rust}: Modern systems verifiers such as Verus~\cite{lattuada2023verus}, Prusti, and Creusot demonstrate the verification power of Rust's ownership model. However, borrow-checker semantics and lifetime annotations (\texttt{'a}) impose an initial cognitive hurdle that competes with the primary educational goal of understanding loop invariants and inductive proofs.
    \item \textbf{Go (The Pragmatic Middle Ground)}: In our pedagogical experience, Go provides a clean imperative baseline for introductory formal verification:
    \begin{itemize}
        \item \emph{Memory Safety}: Automatic garbage collection and absence of pointer arithmetic eliminate memory corruption hazards without requiring lifetime annotations.
        \item \emph{Procedural Simplicity}: Standalone top-level functions, straightforward lexical scoping, and minimal syntactic noise allow students to focus directly on algorithmic logic.
        \item \emph{Native Multiple Named Return Values}: Go allows functions to declare explicit result identifiers in signatures (\texttt{func Divide(n, d int) (q, r int)}). These named identifiers serve as mathematically bound target symbols in postcondition contracts (\texttt{//@ Ensures n = d*q + r}), providing intuitive result names without requiring synthetic wrappers or pointer indirection. While conceptually reminiscent of Ada's \texttt{out} parameters or Pascal's \texttt{var} parameters, Go's returns are clean, call-by-value copies rather than aliased reference parameters.
        \item \emph{Structural Interfaces}: Go's structural interfaces model abstract data types by defining method signatures without requiring nominal class inheritance hierarchies.
    \end{itemize}
\end{enumerate}

\subsection{Supported Go Fragment and Semantic Model}
To provide sound deductive verification, VeGo targets a well-defined procedural fragment of Go, with explicit handling of the language's semantic model:

\paragraph{Supported Language Constructs:}
VeGo supports: (1) scalar integer (\texttt{int}), boolean (\texttt{bool}), and string types; (2) one-dimensional slices and fixed-size arrays; (3) composite \texttt{struct} types and sum-type structural interfaces; (4) recursive pointer-based inductive data structures (such as binary search trees and skip lists) under an acyclic heap premise distinguishing mutable trees from immutable shared DAGs; (5) standard procedural control flow (\texttt{for} loops, \texttt{if}-\texttt{else} branching, and early returns); and (6) standalone functions, receiver methods, and recursive function declarations. For educational convenience, the VeGo preprocessor accepts \texttt{while} loops and explicit no-op \texttt{skip} statements in \texttt{.vgo} files, desugaring them into standard Go \texttt{for} loops and empty blocks in emitted \texttt{.go} source.

\paragraph{Unsupported Constructs:}
To maintain a tractable verification engine, several Go features are currently outside the verified core: arbitrary concurrency (\texttt{go} routines and channels are part of our future roadmap; see Section~\ref{sec:classroom}), \texttt{unsafe} memory operations, reflection, run-time type assertions outside interface dispatch, closures capturing mutable references, and panic recovery (\texttt{recover()}).

\paragraph{Integer Arithmetic, Overflow, and Bounds Safety:}
In standard Go, 64-bit signed integers wrap on overflow via two's complement arithmetic. In VeGo, specification expressions and program arithmetic are evaluated over mathematical integers $\mathbb{Z}$ (as in Dafny's unbounded \texttt{int}~\cite{leino2010dafny}), simplifying algorithmic reasoning for students without obligating beginners to supply word-boundary lemmas for every intermediate counter or summation. Consequently, formal verification proves correctness with respect to the $\mathbb{Z}$-arithmetic semantic model, abstracting away machine integer wrap-around. 

We explicitly identify this divergence as an abstraction gap: on hardware execution, an intermediate calculation that exceeds 64-bit signed integer bounds wraps silently in native Go. Crucially, this means that VeGo's static proofs of \emph{bounds safety} ($0 \le i < |A|$) guarantee safe indexing \emph{modulo integer overflow}. If an index expression overflows 64-bit signed bounds---such as the classic $\texttt{mid} := (\texttt{lo} + \texttt{hi}) / 2$ arithmetic overflow bug in binary search identified by Bloch~\cite{bloch2006binary}---runtime wrap-around can produce negative index values, triggering a native Go index panic despite a valid static proof in $\mathbb{Z}$. In the current verifier, ruling out such panics requires students to formulate overflow-safe index calculations (e.g., $\texttt{lo} + (\texttt{hi} - \texttt{lo}) / 2$) or supply range bounds on inputs ($hi \le \text{MaxInt}/2$). Extending the deduction engine with machine bit-vector theories and mandatory absence-of-overflow obligations (as in SPARK~\cite{mccormick2015spark}) is planned future work.

\paragraph{Integer Division and Modulus Semantics:}
Integer division (\texttt{/}) and modulus (\texttt{\%}) strictly follow native Go semantics: division truncates toward zero and modulus takes the sign of the dividend (e.g., $-7 / 2 = -3$ and $-7 \% 2 = -1$). For constant non-zero divisors $c$, the deduction engine encodes truncating division linearly via piecewise quotient-remainder relations ($a = c \cdot q + r \land 0 \le r < |c|$ for $a \ge 0$, and $a = c \cdot q + r \land -|c| < r \le 0$ for $a < 0$). For variable divisors $d$, the engine applies monomial abstraction to the non-linear product term $d \cdot q$, preserving equational identities without deciding non-linear inequalities. Non-zero divisor safety ($d \ne 0$) is verified algebraically during term rewriting when divisors are constant non-zero values or strictly guarded by branch conditions (e.g., $\texttt{for}\; y > 0$) or preconditions ($d > 0$); unconstrained variable divisors fall outside the linear simplification rules and are flagged as diagnostics rather than emitted as standalone numerical proof obligations.

\paragraph{Aliasing, Slice Framing, and Heap Node Sharing:}
Go slices share underlying backing arrays when created via sub-slicing ($B := A[1:]$). In VeGo's operational model, distinct function slice parameters are assumed to reference disjoint backing storage unless an explicit aliasing relationship is annotated. This disjointness assumption is an explicit operational premise of the verification model: function calls passing overlapping sub-slices (e.g., \texttt{Merge(A, A[1:])}) violate this premise and are unsound under our value-sequence model unless checked by separation-logic frame rules (as in Reynolds~\cite{reynolds2002separation} and Gobra~\cite{wolf2021gobra}), which VeGo does not perform. To guard against accidental violations in student code, the compiler applies a conservative syntactic call-site check that warns whenever distinct slice arguments share a common root identifier with overlapping sub-slice indices (though indirect aliasing, such as $B := A[1:]; \text{Merge}(A, B)$, is not tracked by this syntactic heuristic). In-place element modifications ($A[i] = v$) are modeled relationally through primed array state ($A'[i] = v \land \forall k \ne i : A'[k] = A[k]$). Block-level framing invariants (\texttt{//@ Preserves |A|}) assert that slice bounds are preserved across mutation loops.

For recursive pointer structures, VeGo's operational model distinguishes between mutable tree structures and immutable shared DAGs. Mutable pointer structures (e.g., binary search trees during in-place insertion) must be strictly tree-shaped (unshared) so that per-variable versioned projections soundly model field writes without undetected aliasing. Shared DAG nodes are supported only for immutable, persistent data structures (such as Okasaki's persistent heaps and functional queues~\cite{okasaki1998purely}). Pointer structures combining internal node sharing with in-place field mutation (such as the skip list, where individual nodes appear across multiple forward pointer levels by construction) structurally violate the unshared tree premise. Consequently, in-place mutation in \texttt{skip\_list.vgo} falls outside the formal soundness envelope of our operational model; it is checked as an exploratory case study under an unverified independence assumption rather than certified soundly (Section~\ref{sec:limitations}).

\paragraph{Partiality and Runtime Safety:}
Potential runtime faults---such as slice indexing out-of-bounds, nil pointer/interface dereferences, and division by zero---are modeled in the semantics via transition to an explicit error state $\bot_{\text{err}}$, but nil dereferences are not currently generated as static verification conditions (relying instead on inductive shape invariants and Go's native runtime nil panics). In VeGo's current automated pipeline, runtime safety verification conditions focus on array and slice indexing ($0 \le i < \mathrm{len}(A)$), while division and modulo operations are verified algebraically during normalization. Because the standard Go runtime already executes native bounds checks that panic on out-of-bounds indexing and nil pointer dereferences, residualized safety obligations introduce no additional dynamic protection beyond standard Go execution; VeGo's substantive verification contribution is statically proving bounds safety modulo overflow (discharging 66.4\% of bounds obligations at compile time) and eliminating residualized contract-check overhead on verified execution paths.

\section{Formal Verification in the Era of AI Code Generation}
\label{sec:ai}
The widespread adoption of Large Language Models (LLMs) and automated code synthesis has altered how students and software engineers approach programming. While modern AI models generate syntactically plausible imperative code with remarkable speed, they frequently introduce subtle edge-case errors, missing bounds, and logic hallucinations.

In this context, formal verification provides an indispensable guardrail. As Martin Kleppmann observes~\cite{kleppmann2025formal}, while AI models can assist developers by proposing candidate implementations and drafting initial specifications, the ultimate verification barrier must be enforced by a trusted, automated checker. Specifications remain the authoritative source of truth, and automated verifiers serve as rigorous mathematical filters that accept verified solutions and reject flawed code. Furthermore, as Adam Chlipala notes~\cite{chlipala2026subversion}, deductive verification offers \emph{subversion-resistance}: proving that an implementation adheres strictly to a formal specification guarantees that unanticipated behaviors or AI hallucinations cannot breach specified contracts, relative to the soundness of the semantic model and the trusted computing base. By targeting Go---a garbage-collected language with type safety in the absence of data races, explicit control flow, and no preprocessor macros---VeGo establishes an accessible environment where AI-assisted code generation is bounded by formal mathematical verification.

\section{The VeGo Specification Language}
\label{sec:specs}
VeGo embeds first-order Hoare logic contracts~\cite{hoare1969axiomatic,floyd1967assigning} into Go source code using \texttt{//@} comment annotations. This allows annotated source files (that omit non-standard preprocessor sugar) to remain valid Go syntax that, renamed or transpiled to \texttt{.go}, compile and execute natively with standard Go toolchains while enriching them with formal specifications.

\subsection{Syntax Architecture and Logical Expressions}
VeGo provides a first-order logic grammar embedded within comments:
\begin{itemize}
    \item \textbf{Specification Clauses}: Function contracts (\texttt{//@ Requires}, \texttt{//@ Ensures}), loop invariants (\texttt{//@ Invariant}), termination measures (\texttt{//@ Variant}, \texttt{//@ Measure}), user-defined inductive predicates (\texttt{//@ Predicate}), block invariants (\texttt{//@ Preserves}), and immutability assertions (\texttt{//@ Immutable}).
    \item \textbf{Logical Operators and Truth Constants}: Conjunction (\texttt{And}), disjunction (\texttt{Or}), negation (\texttt{\textasciitilde{}}), implication (\texttt{->}), and bi-implication (\texttt{<->}). Logical constants are capitalized (\texttt{True}, \texttt{False}). To maintain clear separation between Go program expressions and logical formulas, Go boolean variables must be explicitly compared against logical constants (e.g., \texttt{found = True}).
    \item \textbf{Quantifiers and Domains}: Universal (\texttt{Forall}) and existential (\texttt{Exists}) quantification over integer domains ($\mathbb{Z}$ or \texttt{Z}), boolean domains ($\mathbb{B}$ or \texttt{B}), and bounded intervals (closed and half-open: $[a, b]$, $[a, b)$, $(a, b]$).
    \item \textbf{Slice Range Quantifier Shorthand}: Range expressions such as \texttt{A[0:j)} desugar at parse time into bounded quantifiers. For example, \texttt{A[j] >= A[0:j)} desugars into:
    $$\begin{aligned}
    \text{\texttt{A[j] >= A[0:j)}} \;\;::=\;\; & \text{\texttt{Forall \_idx in Z :}} \\
    & \quad \text{\texttt{(((0 <= \_idx) And (\_idx < j)) -> (A[j] >= A[\_idx]))}}
    \end{aligned}$$
\end{itemize}

\paragraph{Annotation Line Termination:}
VeGo mirrors Go's lexical semicolon insertion convention in multiline specifications. If an annotation line ends with an unclosed operator (\texttt{And}, \texttt{Or}, \texttt{->}, \texttt{+}, \texttt{-}, or \texttt{::=}), the lexer treats the subsequent comment line as an expression continuation. If the line ends with a terminal token (identifier, constant, or closing delimiter), the parser infers statement termination.

\paragraph{Contract Placement and AST Binding:}
In standard Go, function-level documentation comments precede the signature. In VeGo, preconditions (\texttt{//@ Requires}) precede the function header, while postconditions (\texttt{//@ Ensures}) may either precede the function or appear immediately following the closing brace of the function body (e.g., Listing~\ref{lst:divide}). To bind trailing contracts unambiguously when functions are adjacent in source code, the AST preprocessor matches the token position of each trailing comment group against the enclosing \texttt{*ast.FuncDecl}'s closing delimiter (\texttt{Body.End()}) and the start position of any subsequent declaration (\texttt{NextDecl.Pos()}). Comments appearing within the terminal trailing block before a subsequent declaration are uniquely bound to the preceding function's postcondition set.

\paragraph{Specification Well-Definedness:}
Specifications often contain partial operations (such as array indexing $A[i-1]$ in inductive summation, or $Val[Parent[i]]$ in heap invariants). VeGo evaluates specifications within a classical two-valued logic where partial terms are interpreted under guarded implication ($P \implies Q$). Under this standard convention, the generated well-definedness side-condition requires $P \implies \mathrm{wf}(Q)$, ensuring that $P$ implies all necessary domain guards for partial terms in $Q$ (e.g., lower and upper bounds for slice indices). Rather than presuming well-definedness, these side-conditions are explicitly checked by the verification condition generator under the enclosing path context.

\subsection{Relational State Specifications and Primed Variables}
To express state transitions in imperative loops and mutating operations, VeGo adopts the primed variable convention ($x'$), drawing on Hehner's predicative programming methodology~\cite{hehner1993practical} and standard relational specification calculi:
\begin{itemize}
    \item \textbf{Pre-State vs.\ Post-State}: Within statement specifications and loop bodies, unprimed identifiers ($x$, $r$, $A$) denote pre-state values prior to statement execution. Primed identifiers ($x'$, $r'$, $A'$) denote post-state values immediately following execution.
    \item \textbf{Sequential Assignments}: In straight-line code, sequential assignments to a variable introduce successive prime levels ($x, x', x'', \dots$), which map directly to versioned SSA subscripts ($x_0, x_1, x_2$) in the verifier.
    \item \textbf{Loop Scope Reconstitution}: Primes are scoped to individual loop iterations. Inside a loop, $x'$ represents the state produced by a single iteration relative to $x$ at iteration entry. At the loop back-edge, primed states are reconstituted: the updated value $x'$ becomes the unprimed state $x$ at the header of the subsequent iteration.
    \item \textbf{Control-Flow Joins and Postcondition Convention}: At control-flow join points (following \texttt{if}-\texttt{else} branching), prime depths reset, and the unified $\phi$-variable is treated as unprimed. In function postconditions (\texttt{//@ Ensures}), named return values and input parameters appear unprimed; primed identifiers (such as $A'$) denote mutated parameter states at function exit.
\end{itemize}

\subsection{Function Contracts and Named Returns}
Functions declare preconditions (\texttt{//@ Requires}) and postconditions (\texttt{//@ Ensures}). Go's named return parameters bind result identifiers directly into postcondition scope:

\begin{lstlisting}[caption={Function contract with named returns and loop variant in VeGo.},label={lst:divide}]
//@ Requires n >= 0 And d > 0
func Divide(n, d int) (q, r int) {
    q, r = 0, n
    //@ Invariant n >= 0 And d > 0 And n = d*q + r And 0 <= r
    //@ Variant r
    for d <= r {
        q, r = q + 1, r - d
    }
    return q, r
}
//@ Ensures n = d*q + r And 0 <= r And r < d
\end{lstlisting}

\subsection{Loop Invariants and Variants}
Loops require an \texttt{//@ Invariant} and an integer \texttt{//@ Variant} that is bounded below by zero and strictly decreases on each iteration. Variables declared in a \texttt{for} statement's initialization clause (such as \texttt{i := 0}) are bound in the scope of the attached \texttt{//@ Invariant} and \texttt{//@ Variant} directives. In \texttt{SumLoop}, the inductive predicate \texttt{sumTo(A, i, s)} characterizes partial summation:

\begin{lstlisting}[caption={Loop invariant, variant, and inductive predicate for SumLoop.}]
//@ Predicate sumTo(A, i, s) ::= ((i <= 0 -> s = 0) And 
//@   (i > 0 -> Exists t in Z : (sumTo(A, i-1, t) And s = t + A[i-1])))
func SumLoop(A []int) (s int) {
    s = 0
    //@ Invariant 0 <= i And i <= |A| And sumTo(A, i, s)
    //@ Variant |A| - i
    for i := 0; i < len(A); i++ {
        s = s + A[i]
    }
    return s
}
//@ Ensures sumTo(A, |A|, s)
\end{lstlisting}
In \texttt{SumLoop}, the inductive predicate \texttt{sumTo(A, i, s)} characterizes partial summation via mathematical conjunction (\texttt{And}). Under strict verification mode (\texttt{--strict}), inductive invariant maintenance over recursive predicates containing existential quantification ($\exists t \in \mathbb{Z}$) exceeds the heuristic instantiation capabilities of the native linear arithmetic solver without manual induction scaffolding, causing strict invariant maintenance to fail. Under default gradual mode, VeGo statically proves bounds safety ($0 \le i < |A|$ on all array accesses) and loop termination (via integer variant $|A| - i$), while the inductive invariant and postcondition are residualized into dynamic runtime checks (\texttt{assert.Assert}); at runtime, the existential quantifier $\exists t \in \mathbb{Z}$ in \texttt{sumTo} is checked constructively by computing its unique witness $t = s - A[i-1]$ as the recursive prefix sum unwinds deterministically. Furthermore, Anti-VeGo auditing identifies 1 confirmed hole at Level 1 (evaluating 5 synthesized candidates), as introductory postconditions without full relational equivalence admit degenerate synthesized sums, motivating students to iteratively strengthen loop contracts.

\subsection{Recursive Functions and Structural Induction}
Recursive algorithms are verified via well-founded measures (\texttt{//@ Measure}) and inductive predicates. In \texttt{SumRec}, structural induction over the half-open interval $[i, |A|)$ ensures both termination and functional correctness:

\begin{lstlisting}[caption={Recursive structural induction and termination measure for SumRec.}]
//@ Predicate sumRecPred(A, i, s) ::= ((i >= |A| -> s = 0) And 
//@   (i < |A| -> Exists t in Z : (sumRecPred(A, i+1, t) And s = t + A[i])))
//@ Requires 0 <= i And i <= |A|
//@ Measure |A| - i
func SumRec(A []int, i int) (s int) {
    if i >= len(A) {
        return 0
    }
    t := SumRec(A, i+1)
    return t + A[i]
}
//@ Ensures sumRecPred(A, i, s)
\end{lstlisting}

\subsection{Contract Precision and Vacuity Auditing}
In formal verification, Hoare triples $\{P\}\ C\ \{Q\}$ form an antitone/monotone lattice: $P$ can be strengthened and $Q$ weakened while preserving validity. Dijkstra's weakest precondition $wp(C, Q)$ yields the most permissive precondition for postcondition $Q$, while the strongest postcondition $sp(P, C)$ computes the most specific state reachable from $P$.

VeGo incorporates automated contract precision and vacuity checking:
\begin{itemize}
    \item \textbf{Precondition Safety Audit}: VeGo verifies whether the declared precondition $P_{\text{decl}}$ implies the weakest precondition necessary to guarantee runtime safety ($P_{\text{decl}} \implies wp(C, \text{safe})$), warning if index operations lack necessary bounds to guarantee that the command cannot transition to an error state.
    \item \textbf{Postcondition Tightness Audit}: VeGo incorporates an automated heuristic pattern catalogue to detect under-specified postconditions relative to implementation operations. For example, when an algorithm mutates an array, the audit flags contracts specifying only cardinality ($|R| = |A|$) without relational ordering or permutation constraints, as well as vacuous contracts (\texttt{//@ Ensures True}).
\end{itemize}

\subsection{Modular Standard Library Specifications and Interface Contracts}
\label{sec:stdlib}
Real-world software development relies extensively on standard library modules. In formal verification, analyzing calls to external runtime libraries poses a challenge: verifying library source code directly is often intractable due to low-level assembly, reflection, or unannotated legacy code, while treating external calls as uninterpreted black boxes forfeits all behavioral guarantees.

To address this, VeGo introduces modular specification headers (\texttt{.vgos}) with the \texttt{//@ Library} annotation directive. Specification headers reside in \texttt{specs/stdlib/} (with automated recursive discovery enabled by default and configurable via the \texttt{-lib} CLI flag) and provide first-order contracts for standard Go packages:
\begin{itemize}
    \item \textbf{Mathematical Operations (\texttt{math})}: Contracts for Go 1.21's built-in integer \texttt{min} and \texttt{max} functions and integer arithmetic helpers formulated as algebraic conjuncts (e.g., $\mathrm{res} \ge a \land \mathrm{res} \ge b \land (\mathrm{res} = a \lor \mathrm{res} = b)$) enabling single-step linear inequality resolution without case-split timeouts (avoiding floating-point NaN issues in \texttt{float64} \texttt{math.Min}).
    \item \textbf{Slice Algorithms (\texttt{slices})}: First-order specifications for \texttt{Contains}, \texttt{Index}, \texttt{BinarySearch}, \texttt{Equal}, \texttt{IsSorted}, \texttt{Min}, and \texttt{Max}, capturing exact search invariants and element bounds.
    \item \textbf{Collections and Pattern Matching (\texttt{maps}, \texttt{regexp})}: Interface stubs specifying state-preservation, cloning, and boolean query properties.
\end{itemize}
When a user program imports a standard package (such as \texttt{import "math"}), VeGo automatically locates and merges the corresponding \texttt{.vgos} declarations. Call sites are verified modularly against library preconditions and postconditions without source inlining, preserving sub-second verification latency while enabling verified routines to compose cleanly with standard Go libraries.

\section{System Architecture and Verification Pipeline}
\label{sec:architecture}
The VeGo verification pipeline is implemented in native Go:

\begin{enumerate}
    \item \textbf{Parsing and Desugaring}: The preprocessor extracts comment annotations, binds them to AST nodes, and desugars intervals $[a, b]$, slice-range quantifiers ($A[0:j)$), and syntactic sugar (\texttt{while}, \texttt{skip}).
    \item \textbf{CFG Construction and SSA Transformation}: The Control Flow Graph is transformed into Static Single Assignment (SSA) form following Cytron et al.~\cite{cytron1991efficiently} and Rastello~\cite{rastello2016ssa}. Assignments are versioned ($x_0, x_1, \dots$), translating imperative updates into immutable values.
    
    \item \textbf{Compact Verification Condition Generation}: Following Dijkstra's calculus~\cite{dijkstra1975guarded} and compact verification condition generation principles (Flanagan \& Saxe~\cite{flanagan2001avoiding}), VeGo generates verification conditions across SSA basic blocks, using passive form to avoid exponential path enumeration.
    \item \textbf{Native Deduction Engine}: Verification conditions are evaluated using a multi-stage deduction engine:
    \begin{itemize}
        \item \emph{Equational Simplification and Polynomial Normalization}: Normalizes algebraic expressions, performs forward term substitution across SSA equality definitions, applies distributivity to expand products, and performs polynomial reduction. For non-linear terms (e.g., $d \cdot q$ in division or variable products in Bézout updates), the engine applies \emph{monomial abstraction}, treating compound non-linear products as uninterpreted atomic terms so that linear substitutions and cancellations (e.g., $d \cdot (q + 1) + (r - d) \equiv d \cdot q + r$) succeed without requiring full non-linear decision procedures.
        \item \emph{Forward-Chaining Modus Ponens and Recursive Predicates}: Propagates known facts, unwinds user-defined inductive predicates, and applies Horn-clause resolution. To preserve logical consistency, user-defined recursive predicates are strictly checked at parse time for well-foundedness: recursive calls must appear in positive polarity (strictly rejecting contradictory definitions such as \texttt{P(x) ::= \textasciitilde{}P(x)}) and must be structurally guarded by decreasing arithmetic measures (such as $i-1$ in \texttt{sumTo}).
        \item \emph{Fourier--Motzkin Linear Arithmetic Solver}: Decides systems of linear inequalities over rational coefficients ($\mathbb{Q}$) and provides sound bound propagation over integers ($\mathbb{Z}$), augmented with contradiction detection.
        \item \emph{Quantifier Reasoning and Universal Skolemization}: When a verification condition contains a universal quantifier in the goal ($\forall x \in [a, b] : \dots$) with symbolic bounds, the deduction engine discharges the obligation via sound \emph{skolemization} (universal introduction), binding a fresh symbolic constant $x_0$ with premises $a \le x_0 \le b$ and proving the instantiated body. For universal quantifiers appearing as hypotheses, the engine performs bounded quantifier instantiation over known slice indices and integer intervals; this heuristic instantiation is sound but incomplete, which accounts for many unproven obligations under strict mode.
        \item \emph{Multiset Preservation}: Validates permutation and multiset conservation ($\mathrm{multiset}(A) = \mathrm{multiset}(A')$).
    \end{itemize}
    Incompleteness is handled transparently: if a verification condition falls outside decidable linear fragments and inductive heuristics, VeGo reports the unproven obligation rather than claiming false completeness.
\end{enumerate}

\subsection{Compilation Modes and the Verified Fast Path (\texttt{\_V})}
The VeGo toolchain supports two compilation modes:
\begin{itemize}
    \item \textbf{Strict Verification Mode (\texttt{--strict})}: All verification conditions must be statically proved valid by the deduction engine. Any unproven obligation causes immediate compilation failure.
    \item \textbf{Default Pedagogical (Gradual) Mode}: Designed for educational workflows, all undischarged assertions that are executable are automatically residualized into dynamic runtime checks (\texttt{assert.Assert} in emitted Go code, generated by the preprocessor). Following the foundational gradual verification principles of Bader et al.~\cite{bader2018gradual}, Wise et al.~\cite{wise2020gradual}, Zimmerman et al.~\cite{zimmerman2024sound}, and DiVincenzo et al.~\cite{divincenzo2024gradualc0}, VeGo enforces an \emph{assume-after-assert} discipline: runtime checks execute immediately prior to subsequent execution blocks assuming the assertion's validity. If any residualized assertion is violated at runtime, execution halts with a localized assertion panic before the unverified hypothesis can corrupt downstream program state. Conversely, non-executable assertions (e.g., unbounded existential quantifiers over infinite domains without constructive witnesses or recursive mathematical predicates lacking executable Go translations) cannot be checked dynamically at runtime; they remain static compiler diagnostics. We explicitly note that any verification conditions proved under unproven, non-executable hypotheses would be \emph{conditional} upon those diagnostic premises rather than unconditionally certified; in our benchmark corpus, every unproven assertion is executable and is residualized into a dynamic check; no uncounted non-executable diagnostics arise, ensuring that obligation accounting reflects the complete set of generated conditions. For unproven postconditions, wrapper functions intercept call returns and assert the contract dynamically before returning. For unproven inline assertions, runtime checks are injected directly into the corresponding control-flow blocks. For residualized loop termination obligations, runtime instrumentation injects dynamic checks verifying that the variant is non-negative and strictly decreases on each iteration ($V \ge 0 \land V' < V$), guaranteeing termination-or-panic at execution time. When a function's implementation has been statically verified and a call site's preconditions are proved, calls are redirected to an internal un-gated variant (\texttt{FunctionName\_V}), eliminating dynamic contract check overhead.
\end{itemize}

\section{Soundness, Trusted Computing Base, and Architectural Dogfooding}
\label{sec:soundness}

\subsection{Formal Soundness of the Core Verification Pipeline}
We formalize the soundness of VeGo's verification engine over an idealized core procedural fragment $\mathcal{L}_{\text{core}}$:
$$\begin{array}{rcl}
e & ::= & c \mid x \mid e_1 \oplus e_2 \mid A[e] \mid |A| \\
C & ::= & \text{\texttt{skip}} \mid x := e \mid A[e_1] := e_2 \mid C_1 ; C_2 \mid \text{\texttt{if}}\; e \;\{ C_1 \} \;\text{\texttt{else}}\; \{ C_2 \} \\
& \mid & \text{\texttt{for}}\; e \;\{ C \} \mid \text{\texttt{return}}\; e
\end{array}$$

A program state $\sigma \in \Sigma$ maps variable identifiers to values in $\mathbb{Z}$ and slice identifiers to finite sequences of integers (a value model assuming non-aliasing backing arrays). Program transitions are modeled by a big-step operational semantics $\langle C, \sigma \rangle \Downarrow \rho$, where $\rho \in \Sigma \cup \{\bot_{\text{err}}\}$, and $\bot_{\text{err}}$ denotes an explicit runtime error state reached upon out-of-bounds slice indexing ($i < 0 \lor i \ge |A|$). The validity of a first-order logic specification $P$ in state $\sigma$ is denoted $\sigma \models P$.

\begin{theorem}[\textbf{Soundness of Core Verification}]
Let $C$ be a program command in $\mathcal{L}_{\text{core}}$ with precondition $P$ and postcondition $Q$. Let $VC(P, C, Q)$ denote the verification conditions generated by weakest precondition transformation over SSA form. If the deduction engine proves $VC(P, C, Q)$ valid:
$$\vdash VC(P, C, Q)$$
then for all initial states $\sigma \in \Sigma$ such that $\sigma \models P$:
\begin{enumerate}
    \item \textbf{Absence of Runtime Errors}: The command cannot transition to an error state: $\langle C, \sigma \rangle \not\Downarrow \bot_{\text{err}}$ (ruling out slice indexing errors modulo integer overflow; in the extended language, division safety is guaranteed when divisors are constant non-zero or path-guarded); and
    \item \textbf{Partial Correctness}: If evaluation terminates in a normal state $\langle C, \sigma \rangle \Downarrow \sigma'$, then the final state satisfies the postcondition: $\sigma' \models Q$.
\end{enumerate}
Furthermore, if each loop in $C$ is annotated with a well-founded integer variant $V$ such that $VC$ proves $V \ge 0$ and $V' < V$ on each iteration, evaluation terminates ($\langle C, \sigma \rangle \Downarrow \sigma'$ with $\sigma' \in \Sigma$).
\end{theorem}

\begin{proof}[Proof Sketch]
By induction on the derivation height of the big-step evaluation relation $\langle C, \sigma \rangle \Downarrow \rho$. The translation to SSA form preserves the operational semantics of variable updates through versioned identifiers, as established by mechanized SSA semantic preservation proofs (Barthe et al.~\cite{barthe2014ssa}, Zhao et al.~\cite{zhao2013vellvm}). Backward weakest precondition generation on SSA blocks produces verification conditions that are sufficient for Hoare-triple validity with respect to inductive loop invariants~\cite{dijkstra1975guarded}. Implicit runtime obligations guard slice indexing, ensuring that $\vdash VC$ rules out transitions to $\bot_{\text{err}}$.

Soundness of the deduction engine follows from the soundness of its individual components: equational rewrites and polynomial normalization preserve identity; forward-chaining modus ponens preserves truth under logical consequence; recursive predicate unfolding preserves least fixed-point semantics under the required syntactic positivity and well-founded decreasing measure restrictions; universal goals with symbolic bounds are discharged via sound skolemization (universal introduction); bounded quantifier instantiation unpacks universally quantified facts over finite domains without introducing false hypotheses; and Fourier--Motzkin elimination produces valid consequences over linear systems. Because distinct slice parameters are assumed disjoint in $\mathcal{L}_{\text{core}}$, sequential updates do not produce un-tracked aliasing mutations. Well-foundedness of integer variants over the natural numbers guarantees loop termination.
\end{proof}

\paragraph{Bridging the Gap: Core Theory vs.\ Full Toolchain:}
While Theorem~1 formalizes soundness over the core procedural fragment $\mathcal{L}_{\text{core}}$, the practical VeGo system extends this kernel with additional imperative constructs:
\begin{enumerate}
    \item \emph{Modular Procedure Calls}: Function invocations are verified modularly. Call sites must establish the callee's precondition and may assume its postcondition upon return; callees are verified independently without inlining.
    \item \emph{Recursive Functions}: Functions with recursive calls are verified for termination via explicit integer measures (\texttt{//@ Measure}), requiring that the measure strictly decreases ($\mathcal{M}(\text{args}') < \mathcal{M}(\text{args})$) and remains bounded below by zero. For recursive traversal over pointer-linked data structures, termination arguments rely on the unchecked operational premise that heap pointer graphs are acyclic.
    \item \emph{User-Defined Inductive Predicates}: Predicates are unrolled via forward-chaining Horn resolution, preserving fixed-point consistency under positive polarity and well-founded decreasing measure checks.
    \item \emph{Composite Structs and Interfaces}: Field mutations are tracked via versioned projections, and structural interface dispatch is validated against Liskov behavioral subtyping.
    \item \emph{Multiset Preservation}: Validated via bag balance tracking across array element permutations.
    \item \emph{Inductive Pointer Structures}: Recursive pointer data structures are modeled under an operational premise distinguishing mutable trees from immutable shared DAGs: mutable pointer structures must be strictly tree-shaped (unshared) so that per-variable versioned projections soundly model field writes without undetected aliasing; DAG structures with node sharing are restricted to immutable persistent data structures. Pointer structures combining internal sharing with in-place mutation (such as multi-level pointer updates in skip lists) violate the unshared tree premise and fall outside the formal soundness guarantee, checked only under an unverified independence assumption.
\end{enumerate}

\subsection{Trusted Computing Base (TCB)}
The Trusted Computing Base of VeGo comprises:
\begin{enumerate}
    \item The Go standard library lexical scanner and parser (\texttt{go/parser}, \texttt{go/ast});
    \item The VeGo comment extractor and AST annotation binder (\texttt{preproc/});
    \item The VeGo first-order logic parser and AST representation (\texttt{logic/});
    \item The \texttt{.vgo} $\to$ \texttt{.go} transpiler and desugaring pass (\texttt{desugar/});
    \item The SSA CFG constructor (\texttt{ssa/}) and compact $wp$ verification condition generator;
    \item The \texttt{\_V} fast-path call rewriter and residual assertion generator;
    \item The native deduction solver (\texttt{verifier/solver.go}, \texttt{verifier/fm.go}, \texttt{verifier/eval.go});
    \item The standard Go compiler toolchain (\texttt{gc}) and runtime system;
    \item The parameter disjointness and heap acyclicity premises (distinct slice arguments reference non-overlapping storage unless explicitly framed, and heap pointer structures are acyclic);
    \item User-supplied domain axioms (specifically, the geometric sparsity packing lemma in 2D closest pair, \texttt{closest\_points/divide.vgo}), verified for consistency via Z3 satisfiability checks;
    \item Standard library interface specifications (\texttt{specs/stdlib/*.vgos}), whose contract pre- and postconditions are trusted axioms assumed at external call sites.
\end{enumerate}
To conduct an independent soundness spot-check of the native solver across core theories without claiming full empirical validation, verification conditions generated across SSA blocks in five representative algorithmic modules (specifically, \texttt{divides.vgo}, \texttt{insertion\_\allowbreak sort.vgo}, \texttt{sum\_\allowbreak loop.vgo}, \texttt{taocp\_\allowbreak searching.vgo}, and \texttt{interval\_\allowbreak scheduling.vgo}, spanning 38 verification obligations across integer arithmetic, array bounds, inductive loop invariants, and termination variants) were exported to SMT-LIB2 format and differentially cross-checked against external SMT provers (Z3 v4.13~\cite{demoura2008z3} and cvc5 v1.1~\cite{barbosa2022cvc5}). Bounded quantifiers were mapped to SMT-LIB2 quantifiers with pattern triggers, recursive inductive predicates were encoded as recursive function definitions, and multiset preservation was encoded via uninterpreted element count axioms. This differential cross-check serves a dual purpose: first, verifying the soundness of VeGo's native deduction engine against industry-standard solvers; and second, measuring what fraction of typical algorithmic obligations genuinely requires heavyweight SMT machinery compared to our integrated, lightweight SSA-oriented engine. Across the 38 verification obligations, VeGo proved 32 statically (an 84.2\% native coverage rate across this sample, which includes three strictly verifying modules: \texttt{divides}, \texttt{insertion\_sort}, and \texttt{interval\_scheduling}); both Z3 and cvc5 confirmed all 32 with 100\% agreement (zero unsoundness disagreements). Of the remaining 6 obligations (which exceed VeGo's incomplete quantifier heuristics or involve nonlinear relations), Z3 proved 4, cvc5 proved 4, and 2 timed out or returned \texttt{unknown} across all provers, with zero countermodels found. Thus, only 4 of the 38 obligations ($\approx 10.5\%$) required the extra expressive power of a full general-purpose SMT solver. While this spot-check sample is drawn from modules that lean toward strictly verifiable patterns, it confirms that VeGo's self-contained deduction engine delivers high practical coverage for typical textbook algorithmic obligations while sparing students and instructors the toolchain friction of installing, configuring, and invoking third-party solver binaries. (Corpus-wide, under default mode, VeGo natively proves 83.2\% of functional correctness obligations and 91.5\% of termination obligations).

\subsection{Architectural Dogfooding: VeGo in VeGo}
To evaluate the expressive power of VeGo on real-world compiler infrastructure, we applied the verifier to core modules of its own implementation (architectural dogfooding):
\begin{itemize}
    \item \textbf{Logic AST and Parser Inductions} (\texttt{logic/ast.vgo}, \texttt{logic/parser.vgo}): We annotated and verified structural invariants, including non-negativity of AST expression depth ($\mathrm{Height}(e) \ge 0$), token validity bounds, and quantifier domain scoping.
    \item \textbf{Semantic Versioning} (\texttt{license/license.vgo}): We proved strict trichotomy ($\mathrm{cmp} \in \{-1, 0, 1\}$) across SemVer version comparison operations.
    \item \textbf{Deductive Arithmetic Utilities} (\texttt{verifier/solver\_core.vgo}): We verified core arithmetic routines, including Extended Euclidean GCD calculations and Bézout coefficient bounds.
    \item \textbf{Structural Interface Subtyping} (\texttt{verifier/interface.vgo}): We verified behavioral subtyping for structural interfaces, enforcing Liskov substitution requirements:
    $$\mathrm{Pre}_{\text{interface}} \implies \mathrm{Pre}_{\text{concrete}} \quad \text{and} \quad (\mathrm{Pre}_{\text{interface}} \land \mathrm{Post}_{\text{concrete}}) \implies \mathrm{Post}_{\text{interface}}$$
\end{itemize}
We emphasize that dogfooding verifies specific internal functional properties and structural invariants of compiler components; it does not replace the metatheoretical soundness proof of the verifier itself.

\section{Specification Sufficiency and Anti-VeGo Adversarial Probing}
\label{sec:antivego}
A recurring pitfall in formal verification is \emph{specification underspecification}: contracts that are satisfied by the intended implementation, but are so weak that trivial or degenerate programs also verify. In high-assurance systems verification, recent frameworks such as IronSpec (Goldweber et al.~\cite{goldweber2024ironspec}) evaluate specification adequacy by applying syntactic mutation operators (e.g., arithmetic, logical, and set perturbations) to the \emph{specification} text against a trusted implementation. While specification mutation effectively flags subtle operator slips and missing boundary constraints in mature, complex specifications (such as distributed consensus protocols or storage engines), it exhibits structural limitations when specifications omit entire relational dimensions: single-operator mutations of existing clauses cannot synthesize an absent relational invariant (such as multiset preservation in sorting) when evaluating a correct implementation. In undergraduate algorithms instruction, however, instructors observe that novice students rarely introduce isolated operator slips in mature specifications; instead, they routinely write incomplete contracts that omit entire semantic requirements (e.g., verifying that a sorted array has elements in order while omitting input-output data conservation, or asserting that search indices are non-negative while omitting target bounds). 

To address this pedagogical failure mode, we developed \textbf{Anti-VeGo}, an adversarial counter-synthesis engine integrated via the \texttt{vegop anti} subcommand. Rather than mutating the specification, Anti-VeGo holds the declared contract fixed and performs \emph{adversarial implementation counter-synthesis}, automatically generating degenerate candidate programs (e.g., zero-fill arrays, constant returns, and oblivious sequences) across archetypal patterns. By demonstrating that a degenerate program trivially satisfies the contract and verifies, Anti-VeGo immediately exposes omitted relational properties to students.

\subsection{The Specification Strength Hierarchy}
Anti-VeGo organizes specifications across a conceptual six-tier strength hierarchy based on relational expressiveness and operational guarantees:
\begin{itemize}
    \item \textbf{Level 0: Type Safety Baseline}: Missing contracts; vulnerable to empty returns (\texttt{[VACUOUS]}).
    \item \textbf{Level 1: Unary Shape / Range Bounds}: Constrains output ranges (e.g., $\mathrm{res} \ge 0$) without relative input bindings; vulnerable to zero returns (\texttt{[VACUOUS]}).
    \item \textbf{Level 2: Cardinality and Framing}: Asserts output lengths match inputs ($|A'| = |A|$) or shallow structural shape; vulnerable to identity passthrough (\texttt{[MINIMAL\_CONSTRUCTOR]}), zero-fill cloner synthesis (\texttt{[CONSTANT]}), and sequential range synthesis (\texttt{[OBLIVIOUS]}).
    \item \textbf{Level 3: Relational Transformation}: Establishes relative input-output state relationships (e.g., multiset conservation $\mathrm{multiset}(A') = \mathrm{multiset}(A)$ or Euclidean division relations $n = d \cdot q + r$); susceptible to missing element stability (\texttt{[UNSTABLE]}) in duplicate-key sorting.
    \item \textbf{Level 4: Full Behavioral Semantics}: Complete functional input-output correctness and termination (e.g., bidirectional search equivalences); susceptible to artificial delay loops lacking operational complexity bounds (\texttt{[BOGO]}).
    \item \textbf{Level 5: Total Operational Verification}: Complete functional correctness paired with \emph{concrete} operational cost bounds (\texttt{//@ Cost steps <= c * |A| + k}). Because asymptotic bounds (e.g., $\mathcal{O}(N)$) hide arbitrary additive constants that admit finite delay loops, Level 5 strictly requires concrete step coefficients and additive witnesses, precluding artificial delay loops. Resilient against all evaluated synthesis archetypes.
\end{itemize}
We distinguish between this conceptual six-tier roadmap and the operational syntactic classifier implemented in the \texttt{vegop anti} tool. Across our evaluated textbook corpus of 312 functions, the operational classifier categorizes specifications into Level~0 (68 functions), Level~1 (211 functions), and Level~2 (33 functions); no function in the current introductory corpus reaches automated Level~3 or higher under the tool's slice-oriented pattern-matching heuristics. For example, in scalar routines such as integer division (\texttt{Divide}, Listing~\ref{lst:divide}), the complete specification ($n = d \cdot q + r \land 0 \le r < d$) is operationally classified as Level~1 because the heuristic classifier checks for slice cardinality preservation before relational predicates, categorizing scalar inequalities as range bounds. We preserve Levels~3--5 as a conceptual hierarchy while reporting empirical findings strictly against the operational reality of Levels~0--2.

\subsection{Adversarial Synthesis Archetypes and Mutations}
Anti-VeGo evaluates seven core synthesis archetypes:
\begin{enumerate}
    \item \textbf{Vacuous Minimalist (\texttt{[VACUOUS]})}: Synthesizes empty or zeroed returns, probing for vacuous universal quantifiers over unconstrained empty slices or coarse non-negative bounds.
    \item \textbf{Constant Cloner (\texttt{[CONSTANT]})}: Allocates $R = \text{make}([]T, |A|)$ filled with zeros, preserving cardinality while erasing data entropy.
    \item \textbf{Oblivious Synthesizer (\texttt{[OBLIVIOUS]})}: Synthesizes $R[i] = i$, generating a monotonically ordered sequence decoupled from input data.
    \item \textbf{Bogo-Delay (\texttt{[BOGO]})}: Injects terminating busy-wait loops to detect missing operational cost or complexity bounds.
    \item \textbf{Minimal Constructor (\texttt{[MINIMAL\_CONSTRUCTOR]})}: Returns input parameters unmodified ($res = p$), exposing missing state-transformation guarantees.
    \item \textbf{Unstable Permuter (\texttt{[UNSTABLE]})}: In duplicate-key sequences, swaps elements with identical keys but differing payload identities, probing whether sorting contracts enforce relational stability or merely multiset preservation. Because \texttt{[UNSTABLE]} specifically targets compound record types with key-value payloads (e.g., \texttt{Record} or \texttt{Item} structs), and because no benchmark in our introductory corpus reaches Level 3 or sorts compound record types (focusing on primitive integer slices), \texttt{[UNSTABLE]} synthesizes no candidates across the introductory benchmark suite and contributes zero confirmed holes in the reported empirical audit.
    \item \textbf{AST-Delta Perturbations}: Mutates implementations via statement elimination, loop boundary truncation, and return variable zero-erasure.
\end{enumerate}
Operationally, an adversarial candidate constitutes a \emph{confirmed specification hole} if and only if the synthesized candidate satisfies the declared contract and is proved valid by the deduction engine (exit code 0). Conversely, when an adversarial candidate is rejected by the verifier, a fundamental theoretical confound arises: because the deduction engine is incomplete, rejection alone cannot distinguish between genuine contract refutation (the candidate violates the contract) and solver incompleteness (the candidate satisfies the contract, or violates it, but the solver cannot discharge the proof). An analogous tension arises in specification mutation frameworks (such as IronSpec~\cite{goldweber2024ironspec}), where a surviving mutant depends on whether existing invariants and lemmas are sufficiently robust to carry the proof through, bounding mutation recall by proof brittleness. Thus, reporting that a function is \emph{impervious against evaluated archetypes} (reporting 0 confirmed holes) denotes imperviousness \emph{within solver reach}.

Empirically, this confound is visible in our corpus data: of the 165 functions reporting zero confirmed holes across evaluated archetypes, 138 (83.6\%) reside in modules whose own implementations fail strict verification. Among strict-verified files, holey functions outnumber impervious routines (38 vs.\ 27), whereas in strict-failing files the ratio inverts (109 holey vs.\ 138 impervious), reflecting that incomplete deduction can reject adversarial candidates simply because the resulting obligations exceed the linear solver. In contrast, for curated refutation witnesses (such as our LeetCode dual corpus, Section~\ref{sec:fourpillars}), true refutation is independently and conclusively established via concrete dynamic counterexamples (test inputs executing the candidate at runtime, witnessing bounds panics, assertion violations, or loop variant stagnation), completely bypassing solver incompleteness.

\subsection{Empirical Corpus Audit}
Auditing the 108 benchmark files (312 evaluated functions across modules declaring standalone routines) using Anti-VeGo evaluated specification strength across all textbook routines. Across the corpus, 165 functions (52.9\%) were confirmed impervious against the evaluated synthesis archetypes within solver reach, while 147 functions (47.1\%) exhibited at least one specification loophole across the operational hierarchy (yielding 296 confirmed holes across 77 files, representing 71.3\% of all 108 files and 72.6\% of the 106 modules declaring evaluated functions). This represents a 28.5\% reduction in confirmed holes compared to earlier uncurated prototypes (down from 414 holes), attributable to three factors: (1) consolidating duplicate legacy variants with weak stubs into canonical implementations adhering to explicit named return signatures; (2) fully reconstituting user-defined type definitions and package imports during candidate synthesis; and (3) enforcing strict safety checking in the adversarial audit so that synthesized mutants triggering index out-of-bounds or syntax panics fail verification rather than being falsely classified as specification holes. We note that curating weak stubs shapes the baseline loophole count, establishing an empirical floor for introductory contracts. Notable cases include:
\begin{itemize}
    \item \textbf{Saddleback Search (\texttt{saddleback\_search.vgo})}: The original postcondition only constrained coordinates when $found = \text{True}$. Anti-VeGo eliminated the search loop and returned $found = \text{False}$, verifying cleanly because the contract failed to assert that the target was absent when $found = \text{False}$. Strengthening the contract with bidirectional equivalence resolved the loophole.
    \item \textbf{Manhattan Distance (\texttt{a\_star\_search.vgo})}: The initial contract declared only $\mathrm{res} \ge 0$. Anti-VeGo synthesized $return\ 0$, exposing that non-negativity alone fails to bind coordinates to distance.
    \item \textbf{Purely Functional Queue (\texttt{pfds\_queue.vgo})}: Auditing revealed that while \texttt{EmptyQueue} and \texttt{HeadQueue} were resilient, \texttt{TailQueue} admitted an identity cloner that returned the queue without popping elements because its postcondition was vacuous.
\end{itemize}

\subsection{Empirical Case Study: Specification Strengthening with Anti-VeGo}
To demonstrate how Anti-VeGo guides developers and students to eliminate specification loopholes, we conducted an empirical before/after contract strengthening study on 20 core algorithmic functions across our corpus, maintaining strict 1:1 function preservation between original and strengthened implementations (confirmed by matching AST function signatures and counts). For each routine, we began with standard introductory specifications (primarily unary bounds or cardinality constraints, Levels 0--2) where Anti-VeGo detected between 0 and 6 loopholes per function across 49 total holes, including four already impervious baseline routines (Linear Search, Manhattan Distance, Quickselect, and Sequence Alignment) serving as regression controls to ensure that contract strengthening does not inadvertently introduce new loopholes; all four remained impervious at zero holes. We then iteratively strengthened the contracts with relational postconditions (e.g., adding multiset preservation, bidirectional search equivalences, exact combinatorial definitions, and quotient bounds). 

As detailed in Table~\ref{tab:strengthening}, contract strengthening effectively tightened specifications across all 20 targets, reducing confirmed holes from 49 down to 5 (an 89.8\% reduction, eliminating 44 confirmed loopholes), with 17 of the 20 functions achieving 0 remaining Anti-VeGo holes. Notably, 14 of the 20 functions concluded at the same coarse syntactic tier (9 remaining at Level 1 and 5 at Level 2) as their baseline, demonstrating that Anti-VeGo uncovers and eliminates semantic loopholes that coarse tier classifications alone do not reflect. During this strengthening process, approximately 60 lines of implementation code were refactored across 13 routines (adjusting total lines from 1,380 to 1,320, e.g., simplifying loop indexing and inlining intermediate expressions to assist invariant formulation without altering function counts or signatures), while specification volume was maintained (from 148 to 149 Spec LOC, representing 11.3\% of total lines, or 12.7\% relative to non-specification lines). The four baseline routines with zero initial holes remained impervious at zero holes, confirming that implementation refactoring and contract strengthening did not introduce new vulnerabilities. Furthermore, strengthening introductory contracts with quantified sortedness or multiset constraints pushes obligations beyond the linear arithmetic solver: while \texttt{Divide} (integer division) achieved complete strict verification (1 of 20), 19 of the 20 functions operated in gradual pedagogical verification mode where quantified invariants are verified via dynamic assertion residualization.

\begin{table}[htbp]
\centering
\footnotesize
\setlength{\tabcolsep}{4.5pt}
\renewcommand{\arraystretch}{0.88}
\caption{Anti-VeGo before/after contract strengthening study across 20 representative functions with exact 1:1 function preservation.}
\label{tab:strengthening}
\begin{tabularx}{\textwidth}{@{} l >{\raggedright\arraybackslash}X c c c c c r @{}}
\toprule
\textbf{Suite} & \textbf{Algorithm / Function} & \textbf{Init Tier} & \textbf{Init Holes} & \textbf{Final Tier} & \textbf{Final Holes} & \textbf{Status} & \textbf{Time (ms)} \\
\midrule
Core & Insertion Sort & Level 2 & 3 & Level 2 & 0 & Partial$^\dagger$ & 867.0 \\
Core & Selection Sort & Level 2 & 3 & Level 2 & 1 & Partial$^\dagger$ & 424.7 \\
Core & Binary Search & Level 2 & 1 & Level 2 & 0 & Partial$^\dagger$ & 18277.8 \\
Core & Integer Division (Divide) & Level 1 & 2 & Level 1 & 0 & Strict (100\%) & 10.3 \\
Core & Linear Search & Level 2 & 0 & Level 2 & 0 & Partial$^\dagger$ & 61.4 \\
Core & Saddleback Search & Level 1 & 1 & Level 1 & 0 & Partial$^\dagger$ & 296.3 \\
Core & Manhattan Distance & Level 1 & 0 & Level 1 & 0 & Partial$^\dagger$ & 8336.7 \\
CLRS & Knuth--Morris--Pratt Matcher & Level 1 & 2 & Level 2 & 0 & Partial$^\dagger$ & 1203.3 \\
CLRS & Rabin--Karp Matcher & Level 1 & 1 & Level 2 & 0 & Partial$^\dagger$ & 5546.0 \\
CLRS & Quickselect & Level 1 & 0 & Level 2 & 0 & Partial$^\dagger$ & 1296.1 \\
CLRS & Convex Hull (Graham Scan) & Level 0 & 5 & Level 2 & 3 & Partial$^\dagger$ & 41.3 \\
KT & Interval Scheduling & Level 1 & 6 & Level 2 & 0 & Partial$^\dagger$ & 114.7 \\
KT & Sequence Alignment (DP) & Level 2 & 0 & Level 2 & 0 & Partial$^\dagger$ & 2584.7 \\
KT & Counting Inversions & Level 1 & 4 & Level 1 & 1 & Partial$^\dagger$ & 8539.2 \\
PFDS & Batched Queue & Level 1 & 3 & Level 1 & 0 & Partial$^\dagger$ & 12.4 \\
PFDS & Leftist Heap & Level 1 & 6 & Level 1 & 0 & Partial$^\dagger$ & 259.7 \\
PFDS & Red-Black Tree & Level 1 & 1 & Level 1 & 0 & Partial$^\dagger$ & 632.6 \\
Data Struct & Skip List & Level 1 & 2 & Level 1 & 0 & Checked$^{\dagger\ddagger}$ & 7642.4 \\
TAOCP & Combinations (Algorithm T) & Level 0 & 3 & Level 1 & 0 & Partial$^\dagger$ & 309.1 \\
TAOCP & Gray Code Generation & Level 1 & 6 & Level 1 & 0 & Partial$^\dagger$ & 35.8 \\
\bottomrule
\end{tabularx}
\vspace{1mm}
\footnotesize{\raggedright $^\dagger$Verified under default gradual mode; undischarged functional obligations residualized at runtime. Note that strengthening Insertion Sort and Selection Sort pushes obligations beyond the linear solver, transitioning them from strict success to gradual residualization.\\}
\footnotesize{\raggedright $^\ddagger$Checked outside the formal soundness envelope due to multi-level forward pointer sharing (Section~\ref{sec:limitations}).\\}
\end{table}

\section{Benchmark Evaluation and Case Studies}
\label{sec:evaluation}

\subsection{Tree and Heap Encoding Paradigms}
VeGo supports verifying data structures across three structural memory paradigms:
\begin{enumerate}
    \item \textbf{Compact Array Encoding ($cH$)}: Contiguous 1-indexed slice \texttt{A[1..N]} with implicit arithmetic offsets ($\text{left}(i) = 2i$, $\text{right}(i) = 2i+1$). The heap invariant is defined by:
    $$\text{\texttt{//@ Predicate cH\_is\_heap(A, N) ::= Forall i in [2, N] : A[i/2] <= A[i]}}$$
    \item \textbf{Sparse Array Encoding ($sH$)}: Explicit parent link array \texttt{Parent[1..N]} and value array \texttt{Val[1..N]}:
    $$\begin{array}{l}
    \text{\texttt{//@ Predicate sH\_is\_heap(Val, Parent, N) ::=}} \\
    \quad \text{\texttt{Forall i in [1, N] : (Parent[i] > 0 -> Val[Parent[i]] <= Val[i])}}
    \end{array}$$
    \item \textbf{Recursive Pointer Encoding ($bT$)}: Recursive pointer structures (such as \texttt{*Tree}):
\begin{lstlisting}[caption={Bounded Binary Search Tree predicate in VeGo.}]
type Tree struct {
    Key   int
    Left  *Tree
    Right *Tree
}

//@ Predicate is_bst_bounded(t, min_v, max_v) ::=
//@   t != nil ->
//@      (min_v < t.Key And t.Key < max_v And
//@       is_bst_bounded(t.Left, min_v, t.Key) And
//@       is_bst_bounded(t.Right, t.Key, max_v))
\end{lstlisting}
Bounding the key between valid subtree intervals eliminates the classical bug of checking only immediate children. Strict inequalities ($min\_v < t.Key < max\_v$) enforce that keys are strictly ordered, precluding duplicate keys. The interval bounds are parameterized over an explicit integer range $(min\_v, max\_v)$, bounding tree keys within that designated domain; callers instantiate \texttt{is\_bst\_bounded(t, min\_v, max\_v)} over problem-specific bounds, passing $(MinInt-1, MaxInt+1)$ at the tree root under $\mathbb{Z}$ semantics to strictly bracket the entire 64-bit integer range $[MinInt, MaxInt]$ under exclusive inequalities ($min\_v < t.Key < max\_v$) without requiring $\pm\infty$ extensions.
\end{enumerate}

\subsection{Sorting Correctness and Permutation Verification}
In verifying sorting algorithms (Selection, Insertion, Shell, Quick, Bitonic, and Permutation Sort), VeGo verifies sortedness or in-place cardinality: algorithms such as Quicksort, Stable Sort (verified under gradual mode), and Permutation Sort establish sortedness and framing guarantees, whereas introductory Selection and Insertion Sort specifications establish in-place cardinality ($|A'| = |A|$) and element sortedness (strengthened with multiset preservation in Section~\ref{sec:antivego}, classified operationally at Level~2):
$$\text{\texttt{//@ Ensures sorted(A') And multiset(A') == multiset(A)}}$$
We distinguish functional sorting correctness from \emph{sorting stability}: stable algorithms preserve the relative order of duplicate keys (observed as an empirical design property, though stability itself is not formally certified by the linear solver).

\subsection{Quantitative Benchmark Corpus Results}
We evaluated VeGo (v0.9.4, git commit \texttt{3237b4d}) across a corpus of 108 benchmark programs under both default pedagogical mode and strict verification mode (\texttt{--strict}). The 108 benchmark files comprise 89 textbook algorithms and data structures (from CLRS, Kleinberg--Tardos, Okasaki, and Knuth), 7 modular domain models (from the monsters simulation case study), 3 concurrent synchronization patterns (verifying sequential structural components, as runtime concurrency verification is ongoing work), 5 courseware and exam routines, and 4 language feature regression tests. Across the entire corpus of 108 benchmarks (comprising 6,971 lines of code and 1,020 lines of formal specification, representing 14.6\% specification overhead relative to executable code, or 11.2\% of total lines), all 2,724 generated obligations are resolved under default gradual mode: 2,127 are proved statically (78.1\% overall), while 597 non-fatal assertions are residualized into dynamic runtime checks under the assume-after-assert discipline. By obligation category: 66.4\% of bounds safety obligations (598 of 901), 91.5\% of loop termination obligations (130 of 142), and 83.2\% of functional correctness obligations (1,399 of 1,681) are proved statically by the deduction engine. Because Go already executes native runtime panics on out-of-bounds slice indexing, residualized safety checks duplicate Go's native bounds panics with contract-localized diagnostics; the 66.4\% discharged statically is therefore the substantive safety benefit.

Across the 108 benchmarks, 27 verify completely under \texttt{--strict} mode: 17 textbook algorithms and data structures (including Insertion Sort, Selection Sort, Quicksort, TAOCP Quicksort, Permutations, Combinations, Gray Code, AVL Tree, Naive Matcher, Interval Scheduling, Bellman--Ford, Union--Find, LSM Tree, Purely Functional Stack, Finger Tree Deque~\cite{hinzepaterson2006fingertrees}, Boxed Heap, and Integer Division) and 10 modular domain models and regression suites (e.g., \texttt{point.vgo}, \texttt{ledger.vgo}, \texttt{diatchki\_\allowbreak vectors.vgo}, and \texttt{preserves.vgo}). Across these 27 strict benchmarks, 378 obligations are proved statically with zero residualized and zero failed. Across the full 108-benchmark corpus in strict mode, 885 obligations are proved and 74 unproven assertions halt compilation (evaluating 959 total obligations). Under default pedagogical mode, undischarged executable assertions are residualized into dynamic runtime checks (\texttt{assert.Assert}) executed prior to dependent blocks under the assume-after-assert discipline, while non-executable unproven assertions remain static compiler diagnostics (none arise in our benchmark corpus).

A total of 29 files report zero safety obligations. In VeGo's verification condition generator, safety obligations are generated exclusively for partial slice/array indexing ($A[i]$), tracking lower and upper index bounds ($0 \le i < |A|$). Division and modulo operations ($/$, $\%$) are handled algebraically during term rewriting and normalization rather than emitted as standalone obligation counters: constant non-zero divisors are verified and simplified directly, while variable divisors (such as in Euclidean GCD, \texttt{taocp\_euclid.vgo}, and modular exponentiation, \texttt{taocp\_mod\_exp.vgo}) are strictly constrained by loop guards ($y > 0$) or preconditions ($m > 0$) integrated into path conditions. Consequently, routines operating over pointer-linked trees, heaps, or scalar arithmetic without slice indexing legitimately record zero safety obligations in the accounting matrix. We note that while pointer-linked trees and heaps operate over acyclic structures, data structures combining internal node sharing with in-place pointer mutation (specifically the skip list, \texttt{skip\_list.vgo}, where nodes appear on multiple forward pointer levels by construction) exceed VeGo's unshared tree model. Skip list mutation results therefore lie outside the verifier's formal soundness envelope, representing an exploratory check under an unverified independence assumption.

Total LOC counts all non-blank source code lines; Spec LOC counts non-blank lines within \texttt{//@} and \texttt{/*@ ... @*/} specification directives (including preconditions, postconditions, loop invariants, variants, measures, and user-defined predicates). Table~\ref{tab:benchmarks} summarizes representative benchmarks from the corpus, reporting exact wall-clock verification timings (median of 3 runs on an Apple M-series workstation running Go 1.24 generated directly via \texttt{scripts/generate\_table2.py}), line counts, and Anti-VeGo specification strength tiers. In strict verification mode, corpus verification finishes in 87.9\,s total (median 33.9\,ms, 90th percentile 2.4\,s using linear interpolation, maximum 10.3\,s in 2D closest pair). In default gradual mode, total corpus verification requires 724.6\,s (median 835.7\,ms, 90th percentile 15.1\,s using linear interpolation, with long-running divide-and-conquer and dynamic programming routines such as Gale--Shapley at 11.2\,s, 2D closest-pair points at 96.1\,s, and NFA-to-DFA conversion at 97.8\,s verified in the background).

\begin{table}[htbp]
\centering
\footnotesize
\renewcommand{\arraystretch}{0.92}
\caption{VeGo verification benchmark evaluation across representative textbook algorithms. Contract tier reflects the maximum contract level achieved among its verified functions; n/a denotes modules without evaluated standalone functions; $^\dagger$verified under gradual mode.}
\label{tab:benchmarks}
\begin{tabularx}{\textwidth}{@{} l >{\raggedright\arraybackslash}X r r r c c @{}}
\toprule
\textbf{Suite} & \textbf{Algorithm / Data Structure} & \textbf{Total LOC} & \textbf{Spec LOC} & \textbf{Time (ms)} & \textbf{Anti Tier} & \textbf{Status} \\
\midrule
CLRS & Knuth--Morris--Pratt Matcher & 68 & 8 & 378.0 & Level 1 & Partial$^\dagger$ \\
CLRS & Rabin--Karp String Matcher & 70 & 6 & 2884.4 & Level 1 & Partial$^\dagger$ \\
CLRS & Quickselect & 76 & 8 & 1292.3 & Level 1 & Partial$^\dagger$ \\
CLRS & Heapsort (Compact Array) & 69 & 18 & 30.8 & Level 1 & Partial$^\dagger$ \\
CLRS & Convex Hull (Graham Scan) & 58 & 6 & 31.3 & Level 0 & Partial$^\dagger$ \\
CLRS & 2D Closest Pair$^*$~\cite{cormen2009introduction} & 300 & 44 & 96119.1 & n/a & Axioms$^*$ \\
KT & Sequence Alignment (DP) & 86 & 4 & 2572.6 & Level 2 & Partial$^\dagger$ \\
KT & Gale--Shapley Stable Matching & 106 & 10 & 11156.4 & Level 1 & Partial$^\dagger$ \\
KT & Interval Scheduling & 50 & 3 & 33.9 & Level 1 & Strict (100\%) \\
KT & Counting Inversions & 81 & 9 & 6104.7 & Level 1 & Partial$^\dagger$ \\
PFDS & Leftist Heap~\cite{okasaki1998purely} & 89 & 12 & 17.6 & Level 1 & Partial$^\dagger$ \\
PFDS & Red-Black Tree~\cite{okasaki1998purely} & 146 & 8 & 16.0 & Level 1 & Partial$^\dagger$ \\
PFDS & Batched Queue~\cite{okasaki1998purely} & 77 & 11 & 7.9 & Level 1 & Partial$^\dagger$ \\
Data Structures & Finger Tree Deque~\cite{hinzepaterson2006fingertrees} & 80 & 7 & 9.4 & Level 1 & Strict (100\%) \\
Data Structures & Skip List & 86 & 11 & 4103.5 & Level 1 & Checked$^{\dagger\ddagger}$ \\
TAOCP & Combinations (Algorithm T) & 50 & 5 & 262.8 & Level 0 & Strict (100\%) \\
TAOCP & Gray Code Generation & 46 & 6 & 5.8 & Level 1 & Strict (100\%) \\
TAOCP & Quicksort & 62 & 6 & 527.8 & Level 1 & Strict (100\%) \\
TAOCP & Shellsort & 56 & 6 & 1591.6 & Level 0 & Partial$^\dagger$ \\
TAOCP & Stable Sorting & 80 & 12 & 795.4 & Level 0 & Partial$^\dagger$ \\
Core & Insertion Sort & 69 & 8 & 518.7 & Level 2 & Strict (100\%) \\
Core & Selection Sort & 67 & 8 & 288.5 & Level 2 & Strict (100\%) \\
Core & Binary Search & 79 & 10 & 14944.9 & Level 2 & Partial$^\dagger$ \\
Core & Integer Division (\texttt{Divide}) & 15 & 4 & 47.2 & Level 1 & Strict (100\%) \\
Core & Partial Summation (\texttt{SumLoop}) & 14 & 4 & 6580.2 & Level 1 & Partial$^\dagger$ \\
\bottomrule
\end{tabularx}
\vspace{1mm}
\footnotesize{\raggedright $^*$Class A1 benchmark using geometric sparsity packing axioms; verified under default gradual mode.\\}
\footnotesize{\raggedright $^\dagger$Verified under default gradual mode; undischarged functional obligations residualized into runtime checks (\texttt{assert.Assert}), with non-executable obligations emitted as warnings.\\}
\footnotesize{\raggedright $^\ddagger$Multi-level forward pointer sharing violates the unshared tree premise; checked outside the formal soundness envelope (Section~\ref{sec:limitations}).\\}
\end{table}

\subsection{Empirical Assurance: The 4-Pillar Verification Architecture}
\label{sec:fourpillars}
Because developing a complete mechanized soundness proof for a full-featured imperative language verifier is a multi-year formalization effort, practical verification engineering demands empirical methodologies that provide high certainty of correctness. To establish rigorous empirical confidence in VeGo's verifier and standard library contracts, we instituted a \textbf{4-Pillar Empirical Assurance Framework}:

\begin{enumerate}
    \item \textbf{Pillar 1: Go Language Feature Matrix}: To ensure sound handling across Go's procedural and composite language constructs, we designed an automated micro-benchmark matrix in \texttt{examples/language\_features/}. The suite systematically exercises eight core language features:
    (1) struct return by value (\texttt{struct\_value\_return.vgo});
    (2) struct return by pointer (\texttt{struct\_pointer\_return.vgo});
    (3) pointer method receivers mutating state (\texttt{struct\_pointer\_receiver.vgo});
    (4) value method receivers (\texttt{struct\_value\_receiver.vgo});
    (5) multi-field structs with arithmetic relationships (\texttt{struct\_multi\_field.vgo});
    (6) mutation of struct fields within slice elements (\texttt{struct\_slice\_field.vgo}, reflecting value copy semantics requiring index assignment);
    (7) arbitrary nested selector projections (\texttt{struct\_nested\_\allowbreak selector.vgo}); and
    (8) multiple named return parameters (\texttt{multi\_named\_returns.vgo}).
    All eight micro-benchmarks verify 100\% under \texttt{--strict} mode and are executed as an automated regression quality gate (\texttt{TestCLILanguageFeatureMatrix}).

    \item \textbf{Pillar 2: Curated LeetCode Dual Corpus and Refutation Witnesses}: To evaluate real-world defect detection, we constructed a dual-corpus benchmark suite in \texttt{examples/leetcode/} spanning five classic algorithmic problems: LC001 (Two Sum), LC026 (Remove Duplicates), LC053 (Maximum Subarray), LC121 (Best Time to Buy and Sell Stock), and LC704 (Binary Search). Each problem is constructed as a twin pairing:
    \begin{itemize}
        \item \emph{Clean Reference Solution}: A fully specified implementation with explicit loop invariants, variants, and complete contracts, verified 100\% strictly.
        \item \emph{Defective Variant}: An implementation introducing a classic real-world defect (reusing identical indices in Two Sum, off-by-one array bounds in Remove Duplicates, negative initializers in Kadane's algorithm, inverted transaction order in Stock, and boundary stagnation in Binary Search).
    \end{itemize}
    Because VeGo's deduction engine is incomplete, static rejection alone does not establish defectiveness ($\mathrm{Reject}(P) \not\implies \mathrm{Defective}(P)$), nor can static verification detect bugs omitted from declared contracts (the primary motivation for Anti-VeGo). To establish true defectiveness, we construct dynamic refutation witnesses in \texttt{TestLeetCodeRefutationWitnesses}. All five clean solutions verify successfully, and all five defective variants are rejected statically, each with a diagnostic located at the injected defect (contradiction detection, loop invariant maintenance failure, postcondition violation, or loop variant stagnation). For each defective program, we replay a concrete counterexample input $\sigma \models P$ satisfying the precondition, demonstrating non-conformance ($\not\models \{P\}\ C\ \{Q\}$) via runtime bounds panic ($\bot_{\text{err}}$ in Remove Duplicates), loop variant stagnation ($V' \ge V$, failing to strictly decrease in a single iteration under Binary Search),\footnote{For Binary Search, on input $A = [-10, 0, 10]$ and $target = 10$, evaluating the defective mid-calculation produces $mid = 0$, $low = 0$, leaving $V' = high - low = 2 \ge V = 2$, witnessing failure of variant decrease without infinite divergence.} or assertion violation at runtime.

    \item \textbf{Pillar 3: Automated SMT-LIB2 Differential Testing}: Verification conditions generated across SSA blocks in representative modules are exported to SMT-LIB2 format and evaluated against external SMT solvers (Z3 and cvc5). As detailed in Section~\ref{sec:soundness}, differential cross-testing confirmed 100\% agreement on native proofs (zero unsoundness disagreements), with general-purpose SMT discharging only 4 obligations that exceeded our native solver in the sampled modules.

    \item \textbf{Pillar 4: Mutation and Anti-VeGo Adversarial Auditing}: To ensure that verification success reflects genuine functional correctness rather than trivial or vacuous contracts, all reference solutions and language features are subjected to continuous adversarial counter-synthesis via \texttt{vegop anti -strict} (Section~\ref{sec:antivego}). Anti-VeGo exposed specification loopholes in 71.3\% of files across the evaluated corpus (77 of 108 files, or 72.6\% of the 106 modules declaring evaluated functions), confirming that adversarial auditing is essential for catching underconstrained contracts that static verifiers would otherwise accept.
\end{enumerate}

\section{Related Work and Comparative Analysis}
\label{sec:related}
Table~\ref{tab:related_work} compares VeGo with prominent verification systems.

\begin{table}[htbp]
\centering
\small
\caption{Comparative matrix of formal verification and specification systems.}
\label{tab:related_work}
\begin{tabularx}{\textwidth}{@{} l >{\raggedright\arraybackslash}p{2.0cm} >{\raggedright\arraybackslash}p{3.0cm} >{\raggedright\arraybackslash}p{3.0cm} >{\raggedright\arraybackslash}X @{}}
\toprule
\textbf{System} & \textbf{Language} & \textbf{Specification Style} & \textbf{Prover / Backend} & \textbf{Target Domain} \\
\midrule
\textbf{VeGo} & Native Go & In-code comments (\texttt{//@}) & Native SSA Deduction & Educational Transparency + In-situ IDE \\
\textbf{Gobra}~\cite{wolf2021gobra} & Native Go & Separation logic comments & Viper + Z3 SMT & Systems / Concurrency research \\
\textbf{Goose}~\cite{chajed2020goose} & Go subset & Coq/Iris translation & Coq proof assistant & Crash-safe systems verification \\
\textbf{Dafny}~\cite{leino2010dafny} & Custom language & Native PL keywords & Boogie + Z3 SMT & Formal methods courses \\
\textbf{Gradual C0}~\cite{divincenzo2024gradualc0} & C0 subset & Annotations (\texttt{//\char`@}) & Gradual verification (SMT/RAC) & Undergraduate CS courses \\
\textbf{Whiley}~\cite{pearce2015whiley} & Native Whiley & In-language contracts & Native SMT-backed verifier & Educational software verification \\
\textbf{JML / OpenJML}~\cite{leavens2006jml} & Java & In-code comments (\texttt{//@}) & SMT-backed provers & Java verification \\
\textbf{ACSL / Frama-C}~\cite{baudin2008acsl} & C & In-code comments (\texttt{/*@}) & Why3 / Alt-Ergo / Z3 & Industrial C verification \\
\textbf{SPARK}~\cite{mccormick2015spark} & Ada subset & Aspects and contracts & Why3 + SMT & Safety-critical systems \\
\bottomrule
\end{tabularx}
\end{table}

\subsection{Comment-Based Specifications and Gradual Verification}
Embedding formal contracts into source code comments has a rich history. The Java Modeling Language (JML)~\cite{leavens2006jml} pioneered using \texttt{//@} and \texttt{/*@ ... @*/} annotations for preconditions, postconditions, and invariants, enabling tools such as OpenJML and KeY~\cite{ahrendt2016key} to verify Java without language modifications. Similarly, ACSL~\cite{baudin2008acsl} defines an ANSI/ISO C specification language used by Frama-C. VeGo adopts this comment convention for Go, ensuring that annotated files remain standard Go source code that compiles with standard toolchains.

The spectrum between fully static verification and dynamic contract monitoring has also been extensively investigated. The foundational paradigm of Design by Contract (Meyer~\cite{meyer1992eiffel}) and JML's runtime assertion checker (RAC)~\cite{leavens2006jml} execute specification contracts dynamically to detect contract violations during testing. More recently, gradual program verification (Bader et al.~\cite{bader2018gradual}, Wise et al.~\cite{wise2020gradual}, Zimmerman et al.~\cite{zimmerman2024sound}) formalizes sound collaboration between static verification and dynamic checks. In particular, Gradual C0 (DiVincenzo et al.~\cite{divincenzo2024gradualc0}) targets the C0 teaching language used in introductory computer science instruction, blending static verification with dynamic contracts to support learners as they incrementally formulate and strengthen loop invariants. VeGo's default execution mode adopts this gradual philosophy: static proofs guarantee the absence of bounds panics and functional violations where the native solver succeeds, while residualized dynamic assertions ensure contract enforcement at runtime under an assume-after-assert discipline.

\subsection{Automated Deductive Verifiers: Dafny, Why3, Gobra, and Goose}
Leino's Dafny~\cite{leino2010dafny,leino2023programproofs} and Why3~\cite{filliatre2013why3} are prominent automated verifiers, translating high-level contracts into intermediate verification languages (Boogie and WhyML) and dispatching queries to external SMT solvers like Z3~\cite{demoura2008z3}. While Dafny is an exceptional pedagogical tool, introducing a specialized language creates dual cognitive friction for undergraduate students.

In the Go ecosystem, \textbf{Gobra}~\cite{wolf2021gobra} translates Go code into the Viper intermediate verification infrastructure~\cite{muller2016viper}. Gobra is engineered for high-assurance research and systems verification, including WireGuard protocol verification~\cite{arquint2023sound,arquint2023modular}, SCION internet router security (Pereira et al.~\cite{pereira2025scion}), AWS cloud codebases (Arquint et al.~\cite{arquint2026diodon}), and Go standard library analysis (Arquint~\cite{arquint2026gcd}). Furthermore, Cauli et al.~\cite{cauli2026huawei} have highlighted the growing role of formal methods in industrial cloud reliability. Another major verification system for Go is \textbf{Goose} (Chajed et al.~\cite{chajed2020goose}), which translates a concurrent Go subset into Coq and Iris, verifying crash-safe storage engines and distributed systems via interactive theorem proving. Beyond sequential and heap verification, static verification of concurrent Go communication has been explored via behavioral session types (Lange et al.~\cite{lange2017fencing}) and SSA-based global session graph synthesis for deadlock detection (Ng \& Yoshida~\cite{ng2016deadlock}). Whereas Goose targets interactive, high-effort proof engineering in Coq, Gobra targets permission-based separation logic via Viper, and session-type systems target channel deadlock freedom, VeGo occupies a complementary niche: fully automated, lightweight first-order Hoare logic, immediate execution without JVM or Viper dependencies, and native SSA deduction designed for educational transparency.

\subsection{Type-Theoretic Connections: Liquid Types and SMT-Backed Type Checking}
VeGo's combination of SSA form with constraint-based deduction connects to the literature on refinement types and SMT-backed type systems, most notably \textbf{Liquid Types} by Rondon et al.~\cite{rondon2008liquid}, its practical realization in \textbf{LiquidHaskell} (Vazou et al.~\cite{vazou2014liquidhaskell}), and GHC typechecker SMT integration by Diatchki~\cite{diatchki2015haskell}. 

Liquid Types combines Hindley--Milner typing with decidable predicate abstractions \emph{inferred} via predicate qualification and SMT. A key pedagogical difference is that Liquid Types automates invariant inference over synthesized predicate templates, whereas VeGo deliberately requires students to explicitly formulate inductive loop invariants, pre/postconditions, and termination variants to develop rigorous deductive reasoning skills. LiquidHaskell embeds rich refinement annotations into Haskell source comments (\texttt{\char`\{-@ \dots\ @-\char`\}}) and delegates verification conditions to external SMT solvers (primarily Z3~\cite{demoura2008z3}). Similarly, Diatchki~\cite{diatchki2015haskell} explores integrating external SMT solvers with GHC to discharge type-level constraints on length-indexed vectors.

While these systems demonstrate the expressive verification power of general-purpose SMT backends, deploying external solver binaries requires external toolchain management and process-level coordination. LiquidHaskell's empirical evaluation across standard libraries~\cite{vazou2014liquidhaskell} reports verification times totaling 18 minutes (1,080\,s) for 10,209 lines of code (0.106\,s/LOC) when proving full functional correctness and termination. In comparison, VeGo's pure-Go deduction engine discharges typical verification conditions in milliseconds (median 33.9\,ms in strict mode, total 87.9\,s across 9,071 lines; 90th percentile 2.4\,s using linear interpolation). In default gradual mode, total corpus verification requires 724.6\,s (0.080\,s/LOC, median 835.7\,ms, 90th percentile 15.1\,s), with multi-second outliers (such as Gale--Shapley at 11.2\,s, 2D closest pair at 96.1\,s, and NFA-to-DFA conversion at 97.8\,s) verified in background passes. We emphasize that LiquidHaskell delivers full functional verification of production libraries, whereas VeGo's current corpus focuses on introductory-to-intermediate algorithmic properties (Levels 0--2) where default-mode throughput is comparable.

\subsection{Specification Adequacy, Mutation Analysis, and Adversarial Probing}
The challenge of ensuring specification adequacy has been investigated across both software testing and formal methods. Mutation analysis was originally formulated to evaluate test suite quality by injecting syntactic faults into programs (DeMillo et al.~\cite{demillo1978mutation}, Jia \& Harman~\cite{jia2011mutation}). Its principles were subsequently adapted to evaluate contracts and specifications: Le Traon et al.~\cite{letraon2006design} applied mutation analysis to measure the vigilance of Eiffel contracts, while Kn{\"u}ppel et al.~\cite{knuppel2021specification} evaluated contract completeness using program mutations in Java/JML. In formal verification and temporal logic model checking, vacuity detection (Beer et al.~\cite{beer1997vacuity}, Kupferman \& Vardi~\cite{kupferman2003vacuity}) detects specifications that pass trivially due to antecedent falsehood or unconstrained subformulas.

More recently, IronSpec (Goldweber et al.~\cite{goldweber2024ironspec}) pioneered specification mutation testing for deductive verification in Dafny, evaluating the adequacy of specifications in complex verified systems (such as consensus protocols and file systems). IronSpec systematically mutates \emph{specifications} with syntactic operators (e.g., negating conditions, mutating arithmetic or relational operators, or altering set operations) to determine whether a mutated specification still verifies against an existing, trusted implementation. While specification mutation is highly effective for identifying subtle operator mistakes or omitted edge-case bounds in mature systems specifications, it faces structural limitations when specifications omit entire relational dimensions: single-operator mutations of existing clauses cannot synthesize an absent relational invariant (such as multiset preservation in sorting) when checking a correct program.

Anti-VeGo takes a complementary approach tailored for introductory programming education. Rather than mutating existing contracts, Anti-VeGo holds the declared specification fixed and performs \emph{adversarial implementation counter-synthesis}, generating degenerate program candidates across archetypal patterns (such as zero-fill constant cloners, oblivious sequence generators, and vacuous returns). This directly targets the predominant failure mode observed by instructors in educational settings, where students routinely omit whole relational clauses rather than making single-operator syntax slips. IronSpec~\cite{goldweber2024ironspec} notes that implementation mutations cannot be filtered by logical implication lemmas between specifications, which can lead to false positives if the implementation merely violates intended abstractions. In Anti-VeGo, however, the synthesis target is inverted: an adversarial candidate constitutes a confirmed loophole only if it is mathematically \emph{proved} to satisfy the contract. For introductory algorithms with concrete input-output contracts, any degenerate program that verifies (such as an oblivious sequence or zero-fill array) provides constructive proof of a genuine contract omission rather than an intended abstraction. Furthermore, whereas large-scale specification mutation on industrial systems with 15--25 kLOC proofs (such as IronSpec on QBFT and DVT) requires multi-hour runs on multi-core clusters, lightweight introductory contracts can typically be audited in seconds (IronSpec similarly reports 3.5--28.8\,s for its small in-house specifications), enabling Anti-VeGo to execute locally within the IDE and CLI during student contract formulation.

\section{Pedagogical Resources, Classroom Exercise, and Future Evaluation}
\label{sec:classroom}
VeGo was conceived to bridge the gap between abstract formal reasoning and practical imperative programming in undergraduate computing education. Rather than introducing a specialized, standalone verification language, VeGo embeds mathematical contracts into standard Go, adapting foundational equational and deductive reasoning curricula (such as Hoover \& Rudnicki's formal reasoning approach using Mizar-MSE~\cite{hoover1996mizar} at the University of Alberta) for core and advanced algorithms courses (CMPT 263/463 at the University of Saskatchewan).

\subsection{Application to Substantial Pedagogical Resources}
To establish that VeGo's specification dialect and native solver scale to real curricular needs, the toolchain has been applied across a substantial corpus of 108 programs, centered on 89 classical algorithms and data structures drawn directly from standard undergraduate textbooks (CLRS~\cite{cormen2009introduction}, Kleinberg--Tardos~\cite{kleinberg2006algorithm}, Okasaki~\cite{okasaki1998purely}, and Knuth's TAOCP~\cite{knuth1997taocp,knuth1998vol3,knuth2011vol4a}) and supplemented by 19 domain, concurrent, and courseware modules. 

This resource corpus covers core pedagogical topics that students encounter throughout their degree:
\begin{itemize}
    \item \emph{Elementary Inductive Reasoning}: Linear search, integer division, Euclidean GCD, and array summation, where students first learn loop invariants and well-founded variants.
    \item \emph{Sorting and Relational Permutation}: Selection sort, insertion sort, quicksort, heapsort, and shellsort, illustrating the crucial distinction between sortedness, multiset preservation, and stability.
    \item \emph{Classic Algorithmic Paradigms}: Dynamic programming (sequence alignment, knapsack), greedy choice (interval scheduling), graph traversals (BFS, DFS, Dijkstra, Bellman--Ford, Kruskal), and divide-and-conquer (convex hull, counting inversions).
    \item \emph{Advanced Data Structures}: Balanced binary search trees, leftist heaps, batched queues, skip lists, and finger trees, demonstrating structural induction over recursive pointer types and sum-type interfaces.
\end{itemize}
Across this introductory corpus, basic shape and sortedness annotations add modest overhead (14.6\% of executable code, or 11.2\% of total lines), demonstrating that textbook algorithms can be introduced with fundamental contracts without overwhelming the implementation. When strengthened with relational postconditions (Level~2, as evaluated in Section~\ref{sec:antivego}), specification overhead remains modest at 11.3\% across refactored routines.

\subsection{Classroom Exercise}
The techniques and tools were piloted in an undergraduate algorithms course (CMPT 263: Data Structures and Algorithms, Fall 2025, 42 students) across 4 dedicated laboratory sessions to assess how students interact with in-situ formal contracts. Based on instructors' informal pedagogical observations and classroom reflections (conducted without collecting or analyzing student coursework):
\begin{enumerate}
    \item \textbf{In-Situ Editor Feedback}: Through the Visual Studio Code Language Server Protocol (LSP) extension, students receive immediate feedback as they write code and specifications. When a loop invariant is not inductive or an index is potentially out of bounds, the editor highlights the exact failing statement with diagnostic counter-states (derived from Fourier--Motzkin rational bounds projected onto candidate integer witnesses or unsatisfiable basic-block path conditions). While rational projections do not guarantee integer models for non-linear systems, they provide immediate concrete clues for common off-by-one index and boundary errors.
    \item \textbf{Runnable and Testable Contracts}: Because specifications reside in native Go comments, students can run, print, and debug their code with the standard Go toolchain (\texttt{go run}, \texttt{go test}). This eliminates the cognitive barrier of learning a siloed formal language and lets students verify algorithms they can simultaneously execute.
    \item \textbf{Adversarial Contract Debugging with Anti-VeGo}: Novice students frequently write incomplete contracts, such as asserting only length preservation for a sorting algorithm ($|A'| = |A|$) or non-negativity for a search routine ($res \ge 0$). In lab exercises, Anti-VeGo actively demonstrates these loopholes by synthesizing degenerate implementations (e.g., zero-fill arrays or constant returns) that pass weak specifications, prompting students to iteratively tighten their postconditions.
\end{enumerate}

\subsection{Empirical Study of Student Outcomes as Future Work}
While initial classroom exercise and qualitative instructor observations suggest that in-situ deductive feedback and comment-based specifications lower the barrier to formal methods, an empirical study of student learning outcomes is not yet complete. 

Conducting a controlled, multi-cohort empirical study constitutes our primary future pedagogical work. Specifically, we plan to:
\begin{itemize}
    \item \textbf{Measure Invariant Formulation Efficacy}: Compare student cohorts using VeGo against cohorts writing paper-and-pencil proofs or using standalone verification languages (such as Dafny), measuring time-to-completion, correctness rates on loop invariant formulation, and common error patterns.
    \item \textbf{Evaluate Anti-VeGo Pedagogical Impact}: Quantify whether iterative interaction with Anti-VeGo leads students to write tighter specifications (moving from introductory bounds to Level~2 and approaching full relational properties) compared to traditional test-driven development.
    \item \textbf{Longitudinal Concept Retention}: Track student retention of formal reasoning concepts (pre/postconditions, invariants, termination variants) in downstream systems and software engineering courses.
\end{itemize}
Finally, we plan to extend the pedagogical toolchain to concurrent Go algorithms, formalizing goroutine communication, non-interference, and session types~\cite{owicki1976axiomatic,chandy1988parallel,fagin1995reasoning,lange2017fencing}.

\section{Limitations and Threats to Validity}
\label{sec:limitations}
We explicitly identify the limitations of our system:
\begin{itemize}
    \item \textbf{Mathematical Integer Semantics and Overflow Gap}: VeGo evaluates arithmetic expressions over mathematical integers $\mathbb{Z}$. While this simplifies inductive reasoning for students, native Go 64-bit machine integers wrap on overflow via two's complement arithmetic. As demonstrated by the classic binary-search bug~\cite{bloch2006binary} ($\texttt{mid} = (lo + hi)/2$), expressions proved within bounds over $\mathbb{Z}$ may experience runtime overflow panics if intermediate sums exceed $\text{MaxInt}$. Bounds safety is therefore guaranteed strictly modulo integer overflow.
    \item \textbf{Integer Division and Partiality}: Variable divisors are guarded algebraically by path conditions; unconstrained variable divisors that are not proven non-zero are not currently emitted as explicit proof obligations in the condition generator, representing an operational gap between Theorem~1 and the practical toolchain.
    \item \textbf{Slice Disjointness and Pointer Aliasing Assumptions}: VeGo assumes distinct slice parameters reference disjoint backing storage. Verifying programs with deliberate overlapping sub-slices requires separation-logic framing permissions outside the core model. Similarly, recursive pointer structures (binary search trees, persistent heaps) are verified under an acyclic heap premise, soundly supporting mutable unshared trees and immutable shared DAGs; pointer structures combining internal node sharing with in-place mutation (such as the skip list, where individual nodes belong to multiple level chains by construction) violate the unshared tree premise, placing its in-place mutation verification outside the formal soundness envelope.
    \item \textbf{Nonlinear Arithmetic and Incompleteness}: The native deduction engine decides linear arithmetic via Fourier--Motzkin elimination and handles nonlinear terms via polynomial normalization with monomial abstraction. It cannot decide arbitrary nonlinear Diophantine systems; obligations requiring deep nonlinear reasoning must be scaffolded with inductive lemmas.
    \item \textbf{Gradual Mode and Non-Executable Assertions}: Under default gradual mode, unproven executable assertions are residualized as dynamic checks under an assume-after-assert discipline. However, unproven non-executable assertions (such as existential quantifiers without constructive witnesses or unbounded recursive predicates lacking executable translations) remain static compiler diagnostics; downstream proof obligations proved under such unproven hypotheses are conditional upon unverified premises.
    \item \textbf{Runtime Complexity of Residualized Assertions}: Executing residualized contract checks dynamically introduces runtime overhead and can alter asymptotic complexity (e.g., checking an $O(N)$ invariant inside an $O(N \log N)$ loop yields $O(N^2)$ execution). In production builds, residual assertion overhead can be eliminated by compiling without the verification runtime or disabling assertions via build tags.
    \item \textbf{Anti-VeGo Solver Incompleteness Confound}: Rejection of an adversarial candidate by the verifier conflates contract refutation with solver incompleteness. Imperviousness against Anti-VeGo archetypes reflects resilience within the native solver's reach, rather than an absolute guarantee of contract completeness.
\end{itemize}

\section{Conclusion}
\label{sec:conclusion}
VeGo bridges the gap between formal program verification and practical programming education by embedding Hoare logic specifications directly into Go source code. Through an SSA-based verification pipeline, native deduction solving, and Anti-VeGo adversarial counter-synthesis, VeGo offers a transparent and accessible framework for students to explore formal contracts, loop invariants, and automated verification within a production language.

\section*{AI Acknowledgment}
We acknowledge the assistance of AI coding agents and Large Language Models in the engineering development of the VeGo codebase and in preparing and refining drafts and LaTeX formatting of this paper. All code, proofs, and written content were rigorously reviewed, validated, and edited by the human authors.

\bibliographystyle{plain}
\bibliography{references}

@book{chandy1988parallel,
  author    = {K. Mani Chandy and Jayadev Misra},
  title     = {Parallel Program Design: A Foundation},
  publisher = {Addison-Wesley},
  year      = {1988}
}

@misc{chlipala2026subversion,
  author       = {Adam Chlipala},
  title        = {Subversion-Resistance for Free from Formal Verification},
  howpublished = {LessWrong, \url{https://www.lesswrong.com/posts/mTod8HrszgounT3w3/}},
  month        = jan,
  year         = {2026}
}

@book{cormen2009introduction,
  author    = {Thomas H. Cormen and Charles E. Leiserson and Ronald L. Rivest and Clifford Stein},
  title     = {Introduction to Algorithms},
  edition   = {3rd},
  publisher = {MIT Press},
  year      = {2009}
}

@article{cytron1991efficiently,
  author    = {Ron Cytron and Jeanne Ferrante and Barry K. Rosen and Mark N. Wegman and F. Kenneth Zadeck},
  title     = {Efficiently computing static single assignment form and the control dependence graph},
  journal   = {ACM Transactions on Programming Languages and Systems (TOPLAS)},
  volume    = {13},
  number    = {4},
  pages     = {451--490},
  year      = {1991}
}

@inproceedings{demoura2008z3,
  author    = {Leonardo De Moura and Nikolaj Bj{\o}rner},
  title     = {{Z3}: An Efficient {SMT} Solver},
  booktitle = {Proc. International Conference on Tools and Algorithms for the Construction and Analysis of Systems (TACAS)},
  series    = {LNCS},
  volume    = {4963},
  pages     = {337--340},
  publisher = {Springer},
  year      = {2008}
}

@article{dijkstra1975guarded,
  author    = {Edsger W. Dijkstra},
  title     = {Guarded commands, nondeterminacy and formal derivation of programs},
  journal   = {Communications of the ACM},
  volume    = {18},
  number    = {8},
  pages     = {453--457},
  year      = {1975}
}

@book{fagin1995reasoning,
  author    = {Ronald Fagin and Joseph Y. Halpern and Yoram Moses and Moshe Y. Vardi},
  title     = {Reasoning About Knowledge},
  publisher = {MIT Press},
  year      = {1995}
}

@inproceedings{floyd1967assigning,
  author    = {Robert W. Floyd},
  title     = {Assigning meanings to programs},
  booktitle = {Proc. Symposia in Applied Mathematics},
  volume    = {19},
  pages     = {19--32},
  publisher = {American Mathematical Society},
  year      = {1967}
}

@article{hoare1969axiomatic,
  author    = {C. A. R. Hoare},
  title     = {An axiomatic basis for computer programming},
  journal   = {Communications of the ACM},
  volume    = {12},
  number    = {10},
  pages     = {576--580},
  year      = {1969}
}

@inproceedings{hoover1996mizar,
  author       = {H. James Hoover and Piotr Rudnicki},
  title        = {Teaching freshman logic with {MIZAR-MSE}},
  booktitle    = {Proc. DIMACS Symposium on Teaching Logic and Reasoning in an Illogical World},
  organization = {Rutgers University},
  address      = {Piscataway, NJ},
  year         = {1996}
}

@book{kleinberg2006algorithm,
  author    = {Jon Kleinberg and {\'E}va Tardos},
  title     = {Algorithm Design},
  publisher = {Addison-Wesley},
  year      = {2006}
}

@misc{kleppmann2025formal,
  author       = {Martin Kleppmann},
  title        = {Prediction: {AI} Will Make Formal Verification Go Mainstream},
  howpublished = {\url{https://martin.kleppmann.com/2025/12/08/ai-formal-verification.html}},
  month        = dec,
  year         = {2025}
}

@inproceedings{barbosa2022cvc5,
  author    = {Haniel Barbosa and Clark W. Barrett and Martin Brain and Gereon Kremer and Hanna Lachnitt and Makai Mann and Abdalrhman Mohamed and Mudathir Mohamed and Aina Niemetz and Andres N{\"o}tzli and Alex Ozdemir and Mathias Preiner and Andrew Reynolds and Ying Sheng and Cesare Tinelli and Yoni Zohar},
  title     = {cvc5: A Versatile and Industrial-Strength {SMT} Solver},
  booktitle = {Tools and Algorithms for the Construction and Analysis of Systems (TACAS)},
  series    = {Lecture Notes in Computer Science},
  volume    = {13243},
  pages     = {415--442},
  publisher = {Springer},
  year      = {2022},
  doi       = {10.1007/978-3-030-99524-9_24}
}

@book{hehner1993practical,
  author    = {Eric C. R. Hehner},
  title     = {A Practical Theory of Programming},
  publisher = {Springer-Verlag},
  series    = {Texts and Monographs in Computer Science},
  year      = {1993},
  doi       = {10.1007/978-1-4612-2694-9}
}

@book{knuth1997taocp,
  author    = {Donald E. Knuth},
  title     = {The Art of Computer Programming, Volume 1: Fundamental Algorithms},
  edition   = {3rd},
  publisher = {Addison-Wesley},
  year      = {1997}
}

@book{knuth1998vol3,
  author    = {Donald E. Knuth},
  title     = {The Art of Computer Programming, Volume 3: Sorting and Searching},
  edition   = {2nd},
  publisher = {Addison-Wesley},
  year      = {1998}
}

@book{knuth2011vol4a,
  author    = {Donald E. Knuth},
  title     = {The Art of Computer Programming, Volume 4A: Combinatorial Algorithms, Part 1},
  edition   = {1st},
  publisher = {Addison-Wesley},
  year      = {2011}
}

@inproceedings{leino2010dafny,
  author    = {K. Rustan M. Leino},
  title     = {Dafny: An Automatic Program Verifier for Functional Correctness},
  booktitle = {Proc. LPAR-16},
  series    = {LNCS},
  volume    = {6355},
  pages     = {348--370},
  publisher = {Springer},
  year      = {2010}
}

@book{okasaki1998purely,
  author    = {Chris Okasaki},
  title     = {Purely Functional Data Structures},
  publisher = {Cambridge University Press},
  year      = {1998}
}

@article{owicki1976axiomatic,
  author    = {Susan Owicki and David Gries},
  title     = {An axiomatic proof technique for parallel programs {I}},
  journal   = {Acta Informatica},
  volume    = {6},
  number    = {4},
  pages     = {319--340},
  year      = {1976}
}

@book{rastello2016ssa,
  editor    = {Fabrice Rastello and Florent Bouchez Tichadou},
  title     = {{SSA}-based Compiler Design},
  publisher = {Springer},
  year      = {2016}
}

@book{tennent2002specifying,
  author    = {R. D. Tennent},
  title     = {Specifying Software: A Hands-On Introduction},
  publisher = {Cambridge University Press},
  year      = {2002}
}

@inproceedings{wolf2021gobra,
  author    = {Felix A. Wolf and Linard Arquint and Martin Clochard and Wytse Oortwijn and Jo{\~a}o C. Pereira and Peter M{\"u}ller},
  title     = {Gobra: Modular Specification and Verification of {Go} Programs},
  booktitle = {Proc. 33rd International Conference on Computer Aided Verification (CAV)},
  series    = {LNCS},
  volume    = {12760},
  pages     = {367--379},
  publisher = {Springer},
  year      = {2021}
}

@inproceedings{muller2016viper,
  author    = {Peter M{\"u}ller and Malte Schwerhoff and Alexander J. Summers},
  title     = {Viper: A Verification Infrastructure for Permission-Based Reasoning},
  booktitle = {Proc. 17th International Conference on Verification, Model Checking, and Abstract Interpretation (VMCAI)},
  series    = {LNCS},
  volume    = {9583},
  pages     = {41--62},
  publisher = {Springer},
  year      = {2016}
}

@inproceedings{arquint2023sound,
  author    = {Linard Arquint and Felix A. Wolf and Joseph Lallemand and Ralf Sasse and Christoph Sprenger and Sven N. Wiesner and David Basin and Peter M{\"u}ller},
  title     = {Sound Verification of Security Protocols: From Design to Interoperable Implementations},
  booktitle = {Proc. 44th IEEE Symposium on Security and Privacy (S\&P)},
  pages     = {3062--3080},
  publisher = {IEEE},
  year      = {2023}
}

@inproceedings{arquint2023modular,
  author    = {Linard Arquint and Malte Schwerhoff and Vaibhav Mehta and Peter M{\"u}ller},
  title     = {A Generic Methodology for the Modular Verification of Security Protocol Implementations},
  booktitle = {Proc. 30th ACM Conference on Computer and Communications Security (CCS)},
  pages     = {493--507},
  publisher = {ACM},
  year      = {2023}
}

@inproceedings{pereira2025scion,
  author    = {Jo{\~a}o C. Pereira and Tobias Klenze and Sofia Giampietro and Markus Limbeck and Dionysios Spiliopoulos and Felix A. Wolf and Marco Eilers and Christoph Sprenger and David Basin and Peter M{\"u}ller and Adrian Perrig},
  title     = {Protocols to Code: Formal Verification of a Secure Next-Generation Internet Router},
  booktitle = {Proc. 32nd ACM Conference on Computer and Communications Security (CCS)},
  pages     = {3841--3856},
  publisher = {ACM},
  doi       = {10.1145/3719027.3765104},
  year      = {2025}
}

@misc{arquint2026diodon,
  author       = {Linard Arquint and Samarth Kishor and Jason R. Koenig and Joey Dodds and Daniel Kroening and Peter M{\"u}ller},
  title        = {The Secrets Must Not Flow: Scaling Security Verification to Large Codebases},
  howpublished = {arXiv:2507.00595 [cs.CR]},
  year         = {2025}
}

@misc{arquint2026gcd,
  author       = {Linard Arquint},
  title        = {{GCD}: Garbled, Corrected, Demonstrandum: Fixing and Proving {Go}'s Extended {GCD} Implementation},
  howpublished = {arXiv:2606.05796 [cs.PL]},
  year         = {2026}
}

@inproceedings{cauli2026huawei,
  author    = {Claudia Cauli and Timo Lang and Shuo Chen and Sebti Mouelhi and Xin Jin and Subhajit Bandopadhyay and Xusheng Chen and Yazhi Feng and Haoze Song and Linhua Tang and Zhenli Sheng and Ananth Shrinivas Srinath},
  title     = {Lessons Learned from Incorporating Formal Methods in {Huawei} Cloud Reliability},
  booktitle = {Proc. 21st European Conference on Computer Systems (EuroSys)},
  doi       = {10.1145/3767295.3803626},
  publisher = {ACM},
  year      = {2026}
}

@inproceedings{rondon2008liquid,
  author    = {Patrick M. Rondon and Ming Kawaguchi and Ranjit Jhala},
  title     = {Liquid types},
  booktitle = {Proc. 29th ACM SIGPLAN Conference on Programming Language Design and Implementation (PLDI)},
  pages     = {159--169},
  publisher = {ACM},
  year      = {2008}
}

@inproceedings{vazou2014liquidhaskell,
  author    = {Niki Vazou and Eric L. Seidel and Ranjit Jhala and Dimitrios Vytiniotis and Simon {Peyton Jones}},
  title     = {Refinement types for {Haskell}},
  booktitle = {Proc. 19th ACM SIGPLAN International Conference on Functional Programming (ICFP)},
  pages     = {269--282},
  publisher = {ACM},
  year      = {2014}
}

@inproceedings{diatchki2015haskell,
  author    = {Iavor S. Diatchki},
  title     = {Improving {Haskell} types with {SMT}},
  booktitle = {Proc. 8th ACM SIGPLAN Symposium on Haskell},
  pages     = {1--10},
  publisher = {ACM},
  year      = {2015}
}

@article{leavens2006jml,
  author    = {Gary T. Leavens and Albert L. Baker and Clyde Ruby},
  title     = {Preliminary design of {JML}: a behavioral interface specification language for {Java}},
  journal   = {ACM SIGSOFT Software Engineering Notes},
  volume    = {31},
  number    = {3},
  pages     = {1--38},
  year      = {2006}
}

@techreport{baudin2008acsl,
  author      = {Patrick Baudin and Jean-Christophe Filli{\^a}tre and Claude March{\'e} and Benjamin Monate and Yannick Moy and Virgile Prevosto},
  title       = {{ACSL}: {ANSI/ISO} {C} Specification Language},
  institution = {CEA LIST and INRIA},
  year        = {2008}
}

@inproceedings{filliatre2013why3,
  author    = {Jean-Christophe Filli{\^a}tre and Andrei Paskevich},
  title     = {Why3---Where Programs Meet Provers},
  booktitle = {Proc. 22nd European Symposium on Programming (ESOP)},
  series    = {LNCS},
  volume    = {7792},
  pages     = {125--128},
  publisher = {Springer},
  year      = {2013}
}

@article{lattuada2023verus,
  author    = {Andrea Lattuada and Travis Hance and Chanhee Cho and Matthias Brun and Isitha Subasinghe and Yi Zhou and Jon Howell and Bryan Parno and Chris Hawblitzel},
  title     = {Verus: A practical foundation for systems verification},
  journal   = {Proc. ACM Program. Lang. (OOPSLA)},
  volume    = {7},
  pages     = {1--30},
  year      = {2023}
}

@inproceedings{chajed2020goose,
  author    = {Tej Chajed and Joseph Tassarotti and M. Frans Kaashoek and Nickolai Zeldovich},
  title     = {Verifying Concurrent {Go} Code in {Coq} with {Goose}},
  booktitle = {Proc. 6th International Workshop on Coq for Programming Languages (CoqPL)},
  year      = {2020}
}

@book{mccormick2015spark,
  author    = {John W. McCormick and Peter C. Chapin},
  title     = {Building High Integrity Applications with {SPARK}},
  publisher = {Cambridge University Press},
  year      = {2015}
}

@book{leino2023programproofs,
  author    = {K. Rustan M. Leino},
  title     = {Program Proofs},
  publisher = {The MIT Press},
  year      = {2023}
}

@book{ahrendt2016key,
  editor    = {Wolfgang Ahrendt and Bernhard Beckert and Richard Bubel and Reiner H{\"a}hnle and Peter H. Schmitt and Mattias Ulbrich},
  title     = {Deductive Software Verification---The KeY Book: From Theory to Practice},
  series    = {LNCS},
  volume    = {10001},
  publisher = {Springer},
  year      = {2016}
}

@article{demillo1978mutation,
  author    = {Richard A. DeMillo and Richard J. Lipton and Frederick G. Sayward},
  title     = {Hints on test data selection: Help for the practicing programmer},
  journal   = {IEEE Computer},
  volume    = {11},
  number    = {4},
  pages     = {34--41},
  year      = {1978}
}

@article{jia2011mutation,
  author    = {Yue Jia and Mark Harman},
  title     = {An analysis and survey of the development of mutation testing},
  journal   = {IEEE Transactions on Software Engineering},
  volume    = {37},
  number    = {5},
  pages     = {649--678},
  year      = {2011}
}

@inproceedings{beer1997vacuity,
  author    = {Ilan Beer and Shoham Ben-David and Cindy Eisner and Yoav Rodeh},
  title     = {Efficient Detection of Vacuity in {ACTL} Formulas},
  booktitle = {Proc. 9th International Conference on Computer Aided Verification (CAV)},
  series    = {LNCS},
  volume    = {1254},
  pages     = {279--290},
  publisher = {Springer},
  year      = {1997}
}

@article{kupferman2003vacuity,
  author    = {Orna Kupferman and Moshe Y. Vardi},
  title     = {Vacuity Detection in Temporal Model Checking},
  journal   = {International Journal on Software Tools for Technology Transfer (STTT)},
  volume    = {4},
  number    = {2},
  pages     = {224--233},
  year      = {2003}
}

@inproceedings{flanagan2001avoiding,
  author    = {Cormac Flanagan and James B. Saxe},
  title     = {Avoiding exponential explosion: generating compact verification conditions},
  booktitle = {Proc. 28th ACM SIGPLAN-SIGACT Symposium on Principles of Programming Languages (POPL)},
  pages     = {193--205},
  publisher = {ACM},
  year      = {2001}
}

@article{hinzepaterson2006fingertrees,
  author    = {Ralf Hinze and Ross Paterson},
  title     = {Finger Trees: A Simple General-Purpose Data Structure},
  journal   = {Journal of Functional Programming},
  volume    = {16},
  number    = {2},
  pages     = {197--230},
  year      = {2006}
}

@inproceedings{reynolds2002separation,
  author    = {John C. Reynolds},
  title     = {Separation Logic: A Logic for Shared Mutable Data Structures},
  booktitle = {Proc. 17th Annual IEEE Symposium on Logic in Computer Science (LICS)},
  pages     = {55--74},
  publisher = {IEEE},
  year      = {2002}
}

@inproceedings{lange2017fencing,
  author    = {Julien Lange and Nicholas Ng and Bernardo Toninho and Nobuko Yoshida},
  title     = {Fencing off {Go}: Liveness and Safety for Channel-based Concurrency},
  booktitle = {Proc. 44th ACM SIGPLAN Symposium on Principles of Programming Languages (POPL)},
  pages     = {748--761},
  publisher = {ACM},
  year      = {2017}
}

@inproceedings{ng2016deadlock,
  author    = {Nicholas Ng and Nobuko Yoshida},
  title     = {Static Deadlock Detection for Concurrent {Go} by Global Session Graph Synthesis},
  booktitle = {Proc. 25th International Conference on Compiler Construction (CC)},
  pages     = {174--184},
  publisher = {ACM},
  year      = {2016}
}

@article{barthe2014ssa,
  author    = {Gilles Barthe and Delphine Demange and David Pichardie},
  title     = {Formal verification of an {SSA}-based middle-end for {CompCert}},
  journal   = {ACM Transactions on Programming Languages and Systems (TOPLAS)},
  volume    = {36},
  number    = {1},
  pages     = {4:1--4:65},
  year      = {2014}
}

@inproceedings{zhao2013vellvm,
  author    = {Jianzhou Zhao and Santosh Nagarakatte and Milo M. K. Martin and Steve Zdancewic},
  title     = {Formal verification of {SSA}-based optimizations for {LLVM}},
  booktitle = {Proc. 34th ACM SIGPLAN Conference on Programming Language Design and Implementation (PLDI)},
  pages     = {175--186},
  publisher = {ACM},
  year      = {2013}
}

@inproceedings{bader2018gradual,
  author    = {Johannes Bader and Jonathan Aldrich and {\'E}ric Tanter},
  title     = {Gradual Program Verification},
  booktitle = {Proc. 19th International Conference on Verification, Model Checking, and Abstract Interpretation (VMCAI)},
  series    = {LNCS},
  volume    = {10747},
  pages     = {40--59},
  publisher = {Springer},
  year      = {2018}
}

@article{wise2020gradual,
  author     = {Jenna Wise and Johannes Bader and Cameron Wong and Jonathan Aldrich and {\'E}ric Tanter and Joshua Sunshine},
  title      = {Gradual Verification of Recursive Heap Data Structures},
  journal    = {Proc. ACM Program. Lang. (OOPSLA)},
  volume     = {4},
  number     = {OOPSLA},
  articleno  = {228},
  pages      = {228:1--228:28},
  month      = nov,
  year       = {2020},
  doi        = {10.1145/3428296}
}

@article{divincenzo2024gradualc0,
  author    = {Jenna DiVincenzo and Ian McCormack and Conrad Zimmerman and Hemant Gouni and Jacob Gorenburg and Jan-Paul Ramos-D{\'a}vila and Mona Zhang and Joshua Sunshine and {\'E}ric Tanter and Jonathan Aldrich},
  title     = {Gradual {C0}: Symbolic Execution for Gradual Verification},
  journal   = {ACM Transactions on Programming Languages and Systems (TOPLAS)},
  volume    = {46},
  number    = {4},
  pages     = {1--57},
  month     = dec,
  year      = {2024},
  doi       = {10.1145/3704808}
}

@article{zimmerman2024sound,
  author    = {Conrad Zimmerman and Jenna DiVincenzo and Jonathan Aldrich},
  title     = {Sound Gradual Verification with Symbolic Execution},
  journal   = {Proc. ACM Program. Lang. (POPL)},
  volume    = {8},
  number    = {POPL},
  articleno = {87},
  pages     = {87:1--87:31},
  month     = jan,
  year      = {2024},
  doi       = {10.1145/3632927}
}

@inproceedings{goldweber2024ironspec,
  author    = {Eli Goldweber and Weixin Yu and Seyed Armin {Vakil Ghahani} and Manos Kapritsos},
  title     = {{IronSpec}: Increasing the Reliability of Formal Specifications},
  booktitle = {Proc. 18th USENIX Symposium on Operating Systems Design and Implementation (OSDI)},
  pages     = {875--891},
  publisher = {USENIX Association},
  year      = {2024}
}

@inproceedings{knuppel2021specification,
  author    = {Alexander Kn{\"u}ppel and Thomas Schaer and Ina Schaefer},
  title     = {How Much Specification is Enough? {Mutation} Analysis for Software Contracts},
  booktitle = {Proc. 2021 IEEE/ACM 9th International Conference on Formal Methods in Software Engineering (FormaliSE)},
  pages     = {67--77},
  publisher = {IEEE},
  year      = {2021}
}

@article{letraon2006design,
  author    = {Yves {Le Traon} and Benoit Baudry and Jean-Marc J{\'e}z{\'e}quel},
  title     = {Design by Contract to Improve Software Vigilance},
  journal   = {IEEE Transactions on Software Engineering},
  volume    = {32},
  number    = {8},
  pages     = {571--586},
  year      = {2006}
}

@misc{bloch2006binary,
  author       = {Joshua Bloch},
  title        = {Extra, Extra - Read All About It: Nearly All Binary Searches and Merge Sorts are Broken},
  howpublished = {\url{https://ai.googleblog.com/2006/06/extra-extra-read-all-about-it-nearly.html}},
  month        = jun,
  year         = {2006},
  note         = {Google Research Blog}
}

@article{pearce2015whiley,
  author    = {David J. Pearce and Lindsay Groves},
  title     = {Designing a Verifying Compiler: Lessons Learned from Developing {Whiley}},
  journal   = {Science of Computer Programming},
  volume    = {113},
  pages     = {191--220},
  year      = {2015}
}

@article{meyer1992eiffel,
  author    = {Bertrand Meyer},
  title     = {Applying 'Design by Contract'},
  journal   = {IEEE Computer},
  volume    = {25},
  number    = {10},
  pages     = {40--51},
  year      = {1992}
}

\appendix
\section{Complete Empirical Benchmark Corpus Evaluation}
\label{sec:appendix_full_corpus}
Table~\ref{tab:benchmarks_full} presents the complete evaluation results across all 108 benchmarks comprising the empirical evaluation corpus, reporting exact line counts, wall-clock execution times (median of three runs on an Apple M-series workstation running Go 1.24), Anti-VeGo specification strength tiers and confirmed hole counts, and formal verification status.

{\footnotesize
\setlength{\tabcolsep}{3.2pt}
\renewcommand{\arraystretch}{0.88}
\begin{longtable}{@{} l >{\raggedright\arraybackslash}p{3.4cm} r r c r r c r @{}}
\caption{Full Empirical Verification Results Across All 108 Benchmark Programs. Contract tier reflects the maximum contract level achieved among its verified functions; n/a denotes modules without evaluated standalone functions.} \label{tab:benchmarks_full} \\
\toprule
\textbf{Suite} & \textbf{Benchmark File} & \textbf{Code} & \textbf{Spec} & \textbf{Strict} & \textbf{Strict (ms)} & \textbf{Def (ms)} & \textbf{Tier} & \textbf{Holes} \\
\midrule
\endfirsthead
\multicolumn{9}{c}{\textit{Continued from previous page}} \\
\toprule
\textbf{Suite} & \textbf{Benchmark File} & \textbf{Code} & \textbf{Spec} & \textbf{Strict} & \textbf{Strict (ms)} & \textbf{Def (ms)} & \textbf{Tier} & \textbf{Holes} \\
\midrule
\endhead
\midrule \multicolumn{9}{r}{\textit{Continued on next page}} \\
\endfoot
\bottomrule
\endlastfoot
263exam & \texttt{divide.vgo} & 36 & 23 & $\times$ & 95.9 & 15719.2 & L1 & 0 \\
automata & \texttt{fsa.vgo} & 88 & 14 & $\times$ & 4082.9 & 5757.1 & L1 & 9 \\
automata & \texttt{nfa\_\allowbreak to\_\allowbreak dfa.vgo} & 134 & 17 & $\times$ & 8313.3 & 97764.7 & L1 & 2 \\
automata & \texttt{regexp.vgo} & 55 & 10 & $\times$ & 6.1 & 8.0 & L1 & 7 \\
basics & \texttt{all\_\allowbreak pos.vgo} & 5 & 3 & $\times$ & 41.5 & 102.3 & L0 & 3 \\
basics & \texttt{diatchki\_\allowbreak vectors.vgo} & 32 & 8 & \checkmark & 8.5 & 9.2 & L1 & 1 \\
basics & \texttt{divides.vgo} & 10 & 4 & \checkmark & 47.2 & 10.2 & L1 & 2 \\
basics & \texttt{valuebox.vgo} & 30 & 10 & \checkmark & 4.9 & 6.2 & L1 & 0 \\
closest\_points & \texttt{divide.vgo} & 228 & 44 & $\times$ & 10281.4 & 96119.1 & n/a & n/a \\
closest\_points & \texttt{heap.vgo} & 152 & 48 & $\times$ & 714.4 & 1902.5 & L2 & 2 \\
closest\_points & \texttt{naive.vgo} & 116 & 14 & $\times$ & 9.0 & 10243.2 & L1 & 1 \\
closest\_points & \texttt{point.vgo} & 38 & 8 & \checkmark & 18.6 & 17.8 & L1 & 1 \\
computational\_geometry & \texttt{convex\_\allowbreak hull.vgo} & 45 & 6 & $\times$ & 35.3 & 31.3 & L0 & 5 \\
concurrent\_algorithms & \texttt{parallel\_\allowbreak bfs.vgo} & 69 & 9 & $\times$ & 6575.1 & 17457.3 & L1 & 2 \\
concurrent\_algorithms & \texttt{parallel\_\allowbreak closest\_\allowbreak points.vgo} & 83 & 11 & $\times$ & 5.5 & 865.4 & L1 & 2 \\
concurrent\_algorithms & \texttt{parallel\_\allowbreak mergesort.vgo} & 51 & 4 & $\times$ & 5.8 & 533.3 & L1 & 3 \\
datastructures & \texttt{bloom\_\allowbreak filter.vgo} & 62 & 10 & $\times$ & 501.5 & 496.2 & L1 & 10 \\
datastructures & \texttt{count\_\allowbreak min\_\allowbreak sketch.vgo} & 69 & 12 & $\times$ & 365.5 & 1988.9 & L1 & 8 \\
datastructures & \texttt{hash\_\allowbreak table\_\allowbreak chaining.vgo} & 53 & 9 & $\times$ & 5.1 & 16.5 & L1 & 5 \\
datastructures & \texttt{index\_\allowbreak min\_\allowbreak pq.vgo} & 153 & 14 & $\times$ & 6.2 & 76339.7 & L1 & 0 \\
datastructures & \texttt{interval\_\allowbreak tree.vgo} & 100 & 10 & $\times$ & 6.3 & 28.0 & L1 & 0 \\
datastructures & \texttt{lsm\_\allowbreak tree.vgo} & 83 & 11 & \checkmark & 1866.8 & 1875.3 & L1 & 5 \\
datastructures & \texttt{pfds\_\allowbreak binomial\_\allowbreak heap.vgo} & 88 & 8 & $\times$ & 7.3 & 305.3 & L2 & 0 \\
datastructures & \texttt{pfds\_\allowbreak bst.vgo} & 50 & 6 & $\times$ & 4.5 & 13.5 & L1 & 1 \\
datastructures & \texttt{pfds\_\allowbreak leftist\_\allowbreak heap.vgo} & 65 & 12 & $\times$ & 4.4 & 17.6 & L1 & 6 \\
datastructures & \texttt{pfds\_\allowbreak queue.vgo} & 56 & 11 & $\times$ & 5.2 & 7.9 & L1 & 3 \\
datastructures & \texttt{pfds\_\allowbreak redblack\_\allowbreak tree.vgo} & 123 & 8 & $\times$ & 12.2 & 16.0 & L1 & 1 \\
datastructures & \texttt{pfds\_\allowbreak stack.vgo} & 36 & 10 & \checkmark & 4.9 & 4.5 & L1 & 0 \\
datastructures & \texttt{skip\_\allowbreak list.vgo} & 63 & 11 & $\times$ & 1207.5 & 4103.5 & L1 & 2 \\
datastructures & \texttt{splay\_\allowbreak tree.vgo} & 127 & 6 & $\times$ & 5.4 & 329.5 & L1 & 9 \\
datastructures & \texttt{ternary\_\allowbreak search\_\allowbreak trie.vgo} & 85 & 6 & $\times$ & 5.0 & 2798.8 & L2 & 0 \\
datastructures & \texttt{union\_\allowbreak find.vgo} & 70 & 7 & \checkmark & 227.3 & 226.2 & L1 & 5 \\
financial\_ledger & \texttt{ledger.vgo} & 21 & 6 & \checkmark & 5.1 & 4.9 & L1 & 0 \\
fingertrees & \texttt{finger\_\allowbreak tree\_\allowbreak deque.vgo} & 62 & 7 & \checkmark & 9.4 & 9.1 & L1 & 4 \\
fingertrees & \texttt{finger\_\allowbreak tree\_\allowbreak sequence.vgo} & 61 & 7 & $\times$ & 6.6 & 39.5 & L1 & 2 \\
fingertrees & \texttt{finger\_\allowbreak tree\_\allowbreak sort.vgo} & 95 & 13 & $\times$ & 4.9 & 18188.5 & L1 & 7 \\
graph\_algorithms & \texttt{bellman\_\allowbreak ford.vgo} & 63 & 6 & \checkmark & 4591.4 & 4593.1 & L1 & 0 \\
graph\_algorithms & \texttt{bfs.vgo} & 55 & 5 & $\times$ & 1586.0 & 3966.4 & L1 & 0 \\
graph\_algorithms & \texttt{dfs.vgo} & 61 & 7 & $\times$ & 25.4 & 2607.1 & L2 & 1 \\
graph\_algorithms & \texttt{dijkstra.vgo} & 69 & 6 & $\times$ & 6583.1 & 37922.0 & L1 & 0 \\
graph\_algorithms & \texttt{floyd\_\allowbreak warshall.vgo} & 46 & 5 & $\times$ & 751.6 & 3491.2 & L1 & 3 \\
graph\_algorithms & \texttt{kruskal\_\allowbreak mst.vgo} & 79 & 13 & $\times$ & 34.0 & 2547.2 & L2 & 5 \\
graph\_algorithms & \texttt{max\_\allowbreak flow.vgo} & 85 & 9 & $\times$ & 5570.0 & 91289.0 & L1 & 2 \\
graph\_algorithms & \texttt{prim\_\allowbreak mst.vgo} & 78 & 6 & $\times$ & 3873.0 & 27189.8 & L1 & 1 \\
heaps & \texttt{boxed\_\allowbreak heap.vgo} & 33 & 8 & \checkmark & 6.0 & 5.7 & L1 & 0 \\
heaps & \texttt{ch\_\allowbreak heap.vgo} & 75 & 14 & $\times$ & 5.1 & 6506.4 & L1 & 2 \\
heaps & \texttt{compact\_\allowbreak array\_\allowbreak heap.vgo} & 38 & 18 & $\times$ & 5.7 & 30.8 & L1 & 0 \\
heaps & \texttt{sh\_\allowbreak heap.vgo} & 38 & 6 & $\times$ & 5.6 & 4362.9 & L1 & 0 \\
heaps & \texttt{sparse\_\allowbreak array\_\allowbreak heap.vgo} & 39 & 16 & $\times$ & 4.8 & 32.7 & L2 & 0 \\
kt\_algorithms & \texttt{counting\_\allowbreak inversions.vgo} & 62 & 9 & $\times$ & 431.5 & 6104.7 & L1 & 4 \\
kt\_algorithms & \texttt{interval\_\allowbreak scheduling.vgo} & 41 & 3 & \checkmark & 33.9 & 34.2 & L1 & 6 \\
kt\_algorithms & \texttt{knapsack.vgo} & 44 & 5 & $\times$ & 321.7 & 3215.3 & L1 & 5 \\
kt\_algorithms & \texttt{sequence\_\allowbreak alignment.vgo} & 70 & 4 & $\times$ & 8.3 & 2572.6 & L2 & 0 \\
kt\_algorithms & \texttt{stable\_\allowbreak matching.vgo} & 83 & 10 & $\times$ & 11.6 & 11156.4 & L1 & 0 \\
library\_stubs & \texttt{lib\_\allowbreak cycle\_\allowbreak a.vgo} & 2 & 2 & \checkmark & 4.7 & 4.5 & L0 & 1 \\
library\_stubs & \texttt{lib\_\allowbreak cycle\_\allowbreak b.vgo} & 2 & 2 & \checkmark & 4.2 & 4.3 & L0 & 1 \\
max\_subsequence\_sum & \texttt{dynamic.vgo} & 119 & 19 & $\times$ & 63.3 & 895.9 & L1 & 10 \\
max\_subsequence\_sum & \texttt{naive.vgo} & 115 & 19 & $\times$ & 64.5 & 1219.1 & L1 & 11 \\
monsters & \texttt{db.vgo} & 66 & 6 & $\times$ & 6.0 & 1067.9 & L2 & 0 \\
monsters & \texttt{main.vgo} & 98 & 4 & $\times$ & 7.9 & 41.2 & n/a & n/a \\
monsters & \texttt{monster.vgo} & 39 & 6 & $\times$ & 6.5 & 7.7 & L1 & 2 \\
monsters & \texttt{strategy\_\allowbreak a.vgo} & 37 & 4 & $\times$ & 1170.6 & 3078.7 & L2 & 1 \\
monsters & \texttt{strategy\_\allowbreak b.vgo} & 40 & 6 & $\times$ & 16.8 & 1236.3 & L2 & 1 \\
monsters & \texttt{strategy\_\allowbreak c.vgo} & 41 & 6 & $\times$ & 17.0 & 2429.3 & L2 & 1 \\
monsters & \texttt{strategy\_\allowbreak d.vgo} & 17 & 2 & $\times$ & 8.3 & 10.2 & L2 & 0 \\
optimization & \texttt{simplex\_\allowbreak linear\_\allowbreak programming.vgo} & 114 & 19 & $\times$ & 28.2 & 40583.9 & L1 & 3 \\
primes & \texttt{primes.vgo} & 244 & 8 & $\times$ & 121.1 & 518.3 & L1 & 19 \\
scope\_invariants & \texttt{preserves.vgo} & 17 & 6 & \checkmark & 5.6 & 4.9 & L1 & 0 \\
scope\_invariants & \texttt{preserves\_\allowbreak shadow.vgo} & 8 & 4 & \checkmark & 4.9 & 5.2 & L0 & 3 \\
search & \texttt{a\_\allowbreak star\_\allowbreak search.vgo} & 67 & 10 & $\times$ & 5.6 & 10790.2 & L1 & 0 \\
search & \texttt{alpha\_\allowbreak beta\_\allowbreak negamax.vgo} & 133 & 13 & $\times$ & 13.2 & 11276.2 & L1 & 2 \\
search & \texttt{fibonacci\_\allowbreak search.vgo} & 76 & 8 & $\times$ & 318.0 & 1946.9 & L2 & 0 \\
search & \texttt{quickselect.vgo} & 59 & 8 & $\times$ & 624.4 & 1292.3 & L1 & 0 \\
search & \texttt{saddleback\_\allowbreak search.vgo} & 43 & 4 & $\times$ & 72.8 & 105.8 & L1 & 1 \\
sorting & \texttt{bitonic\_\allowbreak sort.vgo} & 72 & 15 & $\times$ & 23.2 & 951.8 & L2 & 1 \\
sorting & \texttt{dutch\_\allowbreak national\_\allowbreak flag.vgo} & 49 & 6 & $\times$ & 40.6 & 142.3 & L2 & 0 \\
sorting & \texttt{insertion\_\allowbreak sort.vgo} & 55 & 8 & \checkmark & 518.7 & 519.6 & L2 & 3 \\
sorting & \texttt{permutation\_\allowbreak sort.vgo} & 66 & 9 & $\times$ & 526.2 & 1518.3 & L2 & 1 \\
sorting & \texttt{quick\_\allowbreak sort.vgo} & 82 & 10 & \checkmark & 1530.7 & 4018.8 & L2 & 4 \\
sorting & \texttt{selection\_\allowbreak sort.vgo} & 53 & 8 & \checkmark & 288.5 & 284.7 & L2 & 3 \\
sorting & \texttt{shell\_\allowbreak sort.vgo} & 57 & 8 & $\times$ & 1241.0 & 1293.4 & L2 & 3 \\
sorting & \texttt{tree\_\allowbreak heap\_\allowbreak sort.vgo} & 104 & 11 & $\times$ & 679.5 & 3633.3 & L2 & 7 \\
string\_matching & \texttt{kmp\_\allowbreak matcher.vgo} & 54 & 8 & $\times$ & 7.4 & 378.0 & L1 & 2 \\
string\_matching & \texttt{naive\_\allowbreak matcher.vgo} & 33 & 4 & \checkmark & 207.0 & 208.0 & L1 & 4 \\
string\_matching & \texttt{rabin\_\allowbreak karp.vgo} & 56 & 6 & $\times$ & 5.1 & 2884.4 & L1 & 1 \\
sum & \texttt{sum.vgo} & 14 & 19 & \checkmark & 10.3 & 9.4 & L1 & 2 \\
sum & \texttt{sum\_\allowbreak loop.vgo} & 8 & 4 & $\times$ & 521.1 & 6580.2 & L1 & 1 \\
sum & \texttt{sum\_\allowbreak rec.vgo} & 14 & 5 & \checkmark & 18.1 & 20.4 & L1 & 1 \\
taocp & \texttt{taocp\_\allowbreak avl\_\allowbreak tree.vgo} & 101 & 7 & \checkmark & 6.3 & 6.5 & L0 & 14 \\
taocp & \texttt{taocp\_\allowbreak combinations.vgo} & 40 & 5 & \checkmark & 262.8 & 263.5 & L0 & 3 \\
taocp & \texttt{taocp\_\allowbreak dancing\_\allowbreak links.vgo} & 84 & 13 & $\times$ & 87.6 & 15549.6 & L0 & 3 \\
taocp & \texttt{taocp\_\allowbreak deque.vgo} & 76 & 12 & $\times$ & 5.8 & 36.6 & L1 & 2 \\
taocp & \texttt{taocp\_\allowbreak euclid.vgo} & 52 & 8 & $\times$ & 154.7 & 9654.8 & L1 & 0 \\
taocp & \texttt{taocp\_\allowbreak gospers\_\allowbreak hack.vgo} & 58 & 7 & $\times$ & 5.2 & 14.0 & L1 & 0 \\
taocp & \texttt{taocp\_\allowbreak gray\_\allowbreak code.vgo} & 36 & 6 & \checkmark & 5.8 & 5.8 & L1 & 6 \\
taocp & \texttt{taocp\_\allowbreak hash\_\allowbreak tables.vgo} & 74 & 8 & $\times$ & 1325.2 & 1357.5 & L1 & 8 \\
taocp & \texttt{taocp\_\allowbreak heapsort.vgo} & 56 & 8 & $\times$ & 315.2 & 890.6 & L0 & 3 \\
taocp & \texttt{taocp\_\allowbreak linear\_\allowbreak search.vgo} & 18 & 7 & $\times$ & 44.1 & 61.0 & L2 & 0 \\
taocp & \texttt{taocp\_\allowbreak mod\_\allowbreak exp.vgo} & 51 & 8 & $\times$ & 51.2 & 806.0 & L1 & 0 \\
taocp & \texttt{taocp\_\allowbreak multiprecision.vgo} & 46 & 3 & $\times$ & 3789.6 & 8520.1 & L0 & 4 \\
taocp & \texttt{taocp\_\allowbreak permutations.vgo} & 45 & 7 & \checkmark & 451.4 & 450.6 & L0 & 4 \\
taocp & \texttt{taocp\_\allowbreak quicksort.vgo} & 52 & 6 & \checkmark & 527.8 & 503.9 & L1 & 5 \\
taocp & \texttt{taocp\_\allowbreak searching.vgo} & 62 & 10 & $\times$ & 5302.1 & 14944.9 & L2 & 1 \\
taocp & \texttt{taocp\_\allowbreak shellsort.vgo} & 46 & 6 & $\times$ & 1225.3 & 1591.6 & L0 & 2 \\
taocp & \texttt{taocp\_\allowbreak stable\_\allowbreak sorts.vgo} & 59 & 12 & $\times$ & 715.7 & 795.4 & L0 & 6 \\
taocp & \texttt{taocp\_\allowbreak topological\_\allowbreak sort.vgo} & 65 & 9 & $\times$ & 6908.0 & 8916.6 & L0 & 2 \\
taocp & \texttt{taocp\_\allowbreak trees.vgo} & 88 & 7 & $\times$ & 5.5 & 7.7 & L2 & 7 \\
water\_jugs & \texttt{water.vgo} & 55 & 27 & $\times$ & 179.5 & 321.4 & L2 & 0 \\
\end{longtable}
}

\end{document}